\documentclass[12pt, reqno, oneside,amsart]{article}
\usepackage[
left=1.3in,
right=1.3in,
top=1.in,
bottom=1.in,
includeheadfoot,
footskip=0.7in
						]{geometry} 
\usepackage{amsmath}

\usepackage{cite}
\usepackage{graphicx}
\usepackage{amsfonts}
\usepackage{amssymb}
\input epsf

\usepackage[shortlabels]{enumitem}

\usepackage[makeroom]{cancel}	

\usepackage{framed}

\usepackage{braket}

 \usepackage[utf8]{inputenc}

\usepackage{lipsum}

\usepackage[titletoc]{appendix}

\usepackage{enumitem}

\usepackage[dvipsnames]{xcolor}

\usepackage{hyperref}
\hypersetup{
    colorlinks=true,
    anchorcolor=red,
    urlcolor=magenta,
    citecolor=red,
    linkcolor=cyan}
    
\usepackage{simplewick}    

\usepackage{empheq}

\usepackage{mathrsfs}  

\usepackage{dsfont}

\usepackage{wrapfig}

\usepackage{stmaryrd}

\usepackage{subfigure}

\usepackage{amsmath}

\usepackage{slashed}

\makeatletter

\def\env@sqcases{%
  \let\@ifnextchar\new@ifnextchar
  \left\lbrack
  \def\arraystretch{1.2}%
  \array{@{}l@{\quad}l@{}}%
}
\makeatother

\numberwithin{equation}{section}

\usepackage{etoolbox}
\usepackage{titlesec}
\usepackage{titletoc}
\usepackage{titletoc}

\titleformat{\section} 
	{
	\filcenter
	\normalsize
	\scshape
	} 
{\thesection.}       
{0.3em}                  
{}

\titlespacing{\section}
{0pt}                  
{20pt}                  
{13pt}                  

\titleformat{\subsection} 
{
\bfseries
} 
{\thesubsection}       
{0.3em}                  
{}                     

\titlespacing{\subsection}
{0pt}                  
{20pt}                  
{5pt}                  

\titleformat{\subsubsection} 
[hang]
{
\normalsize
\bfseries
} 
{\thesubsubsection.}    
{0.3em}                
{}                     

\titlespacing{\subsubsection}
{0pt}                  
{25pt}		        
{15pt}                  

\def\a{\alpha}
\def\b{\beta}
\def\s{\sigma}

\def\la{\lambda}
\def\La{\Lambda}

\def\ga{\gamma}

\def\e{\epsilon}

\def\rm{\mathrm}
\def\cal{\mathcal}
\def\scr{\mathscr}

\def\pa{\partial}

\def\be{\begin{equation}}
\def\ee{\end{equation}}
\def\br{\begin{eqnarray}}

\def\er{\end{eqnarray}}
\def\bsub{\begin{subequations}}
\def\esub{\end{subequations}}

\def\R{\mathrm{R}}

\def\Oint{O^{(\rm{int})}_{[2]}}

\def\Oincov{O^{(\rm{int})}}

\def\p{\partial}

\def\bphi{\tilde\phi}

\title{Dynamics of R-neutral Ramond fields in the D1-D5 SCFT}

\date{} 

\usepackage{authblk}

\author[1]{\normalsize A.~A. Lima\thanks{andrealves.fis@gmail.com}}
\author[1]{\normalsize G.~M. Sotkov\thanks{gsotkov@gmail.com}}
\author[2]{\normalsize M. Stanishkov\thanks{marian@inrne.bas.bg}}

\affil[1]{\textit{\footnotesize Department of Physics, Federal University of Esp\'irito Santo, 29075-900, Vit\'oria, Brazil}}
\affil[2]{\textit{\footnotesize Institute for Nuclear Research and Nuclear Energy, Bulgarian Academy of Sciences, 1784 Sofia, Bulgaria}}

\begin{document}

\begin{titlepage}

\maketitle

\begin{abstract}

We describe the effect of the marginal deformation of the $\cal N = (4, 4)$ superconformal $(T^4)^N /S_N$ orbifold theory on a doublet of R-neutral twisted  Ramond fields, in the  large-$N$  approximation.
Our analysis of their dynamics explores  the explicit analytic form of the genus-zero four-point function involving two R-neutral Ramond fields and two deformation operators.
We compute this correlation function with two different approaches: the Lunin-Mathur path-integral technique and the stress-tensor method. From its short distance limits, we extract the OPE structure constants and the scaling dimensions of non-BPS fields appearing in the fusion.
 In the  deformed CFT, at  second order in the deformation parameter,
 the two-point function of the $n$-twisted Ramond fields is UV-divergent.
We perform an appropriate regularization, together with a renormalization  of the undeformed fields, obtaining finite, well-defined corrections to their two-point functions and their bare conformal weights, for $n < N$. The fields with maximal twist $n=N$ remain protected from renormalization, with vanishing anomalous dimensions.

{\footnotesize 
\bigskip
\noindent
\textbf{Keywords:}

\noindent
Symmetric product orbifold of $\mathcal {N}=4$ SCFT, marginal deformations, twisted Ramond fields, correlation functions, anomalous dimensions.
}

\end{abstract}

\pagenumbering{gobble}

\end{titlepage}

\pagenumbering{arabic}

\tableofcontents

\section{Introduction} \label{sec:Introduction}%

The powerful machinery of two-dimensional conformal field theories makes the  AdS$_3$/CFT$_2$ duality a special framework for the exploration of the holographic correspondence.
One of its most effective applications is in the description of the near-horizon limit of the geometry created by the bound state of a large number of D1 and D5 branes in Type-IIB supergravity \cite{Maldacena:1997re}, where one obtains an AdS$_3 \times S^3$ background compactified on $T^4$.%
\footnote{%
It is possible to compactify on K3 instead, but we will not consider this case.}
The corresponding  $\cal N = (4,4)$ supersymmetric CFT lies in the moduli space of the free $(T^4)^N/S_N$ orbifold  CFT, where $S_N$ is the symmetric group of $N = N_1 N_5$ elements, with $N_1,N_5$ being the large numbers of D1- and D5-branes \cite{Seiberg:1999xz,Larsen:1999uk}.
The relation between the gravitational solutions and the CFT states has played a  crucial role in counting black hole degrees of freedom \cite{Strominger:1996sh,Maldacena:1998bw,Seiberg:1999xz,Larsen:1999uk,Maldacena:1997re,David:2002wn}, and in understanding the microscopic structure of fuzzballs \cite{Lunin:2001jy,Mathur:2005zp,Kanitscheider:2007wq,Kanitscheider:2006zf,Skenderis:2008qn,Bena:2010gg,Giusto:2012yz,Giusto:2013bda,Bena:2015bea,Bena:2016ypk,Bena:2018mpb,Warner:2019jll}.
The free orbifold CFT has also provided an important example where the exact realization of AdS$_3$/CFT$_2$ is within reach
\cite{Galliani:2017jlg,Giusto:2019qig,Eberhardt:2019qcl,Eberhardt:2018ouy,Dei:2019iym,Giusto:2020mup,Gaberdiel:2020ycd}.

The twisted sectors of the CFT correspond to single-cycle permutations of length $n \leq N$ in $S_N$. They form the fundamental blocks of the total Hilbert space, as a generic permutation in $S_N$ can be uniquely decomposed in products of cycles of different lengths.
The correlation functions of twisted operators have monodromies determined by how the cyclic permutations combine with each other, making the orbifold CFT non-trivial even at the free point of moduli space.
The deformation operator which drives the CFT towards the SUGRA region is itself twisted. Specifically, the deformed action is \cite{Avery:2010er,David:2002wn,Seiberg:1999xz}
\be
S_{\rm{int}} (\lambda)=S_{\rm{free}} + \la \int \! d^2 z \, \Oint (z,\bar z), \label{def-cft}
\ee
where $\la$ is a dimensionless coupling constant and $\Oint (z,\bar z)$ a scalar modulus marginal operator with twist 2.
This twist introduces a further complication, as it changes the lengths of the cycles of other twisted operators with which $\Oint$ interacts, either by joining previously disjoint cycles or splitting a cycle in two.
The effect of the deformation operator on states of the orbifold have been investigated in several works  \cite{Avery:2010er,Avery:2010hs,Carson:2014ena,Carson:2015ohj,Guo:2020gxm,Hampton:2018ygz,Guo:2019ady,Guo:2019pzk,Burrington:2012yq,Burrington:2017jhh,Keller:2019yrr,Carson:2016uwf,Keller:2019suk}, using different methods.

The purpose of the present paper is to discuss the effect of $\Oint$ on the R-neutral Ramond ground states of the $n$-twisted sector,
extending the analysis of R-charged Ramond fields $R^{\pm}_{n}$ done in \cite{Lima:2020boh,Lima:2020kek}, and thus forming a complete picture for all single-cycle Ramond ground states in the deformed CFT (\ref{def-cft}).
The R-neutral Ramond fields, which we denote by $R^{0\pm}_{n}$,
are single-cycle $n$-twisted generalizations of spin fields.
In the `seed' CFT, which has $c = 6$ and target space $T^4$, there are 4 holomorphic Ramond ground states with conformal weights $h^\R = \frac{c}{24}$, distinguished by their holomorphic and anti-holomorphic R-charges, and by their charges under the ``internal'' $\rm{SO}(4) = \rm{SU}(2)_1 \times \rm{SU}(2)_2$ group. The Ramond fields $R^{0\pm}_n$ considered here have zero R-charges, and form a doublet under the internal current of $\rm{SU}(2)_2$; they also have $h^\R_n = \frac1{4} n$, as the CFT in the $n$-twisted sector has $c = 6n$.

Ramond ground states are fundamental pieces of the orbifold CFT.
First and foremost they are, as said, the ``spin fields of the twisted sectors''. Spin fields have the responsibility of creating the anti-periodic boundary conditions of fermions in the Ramond sector of a (non-orbifolded) CFT defined on the complex plane.
The $n$-twisted Ramond fields have the equivalent effect in the sectors of the orbifold, but they are even more fundamental because subtleties of the periodicity of the twisted fermions are such that, for even $n$, one is \emph{forced} to apply either one of $R^{0\pm}_{n}$ or $R^\pm_{n}$ on the vacuum, for the theory to be well-defined  \cite{Avery:2009xr,Lunin:2001pw}.
In this sense, when fermions are involved, the Ramond fields are as basic as the bare twist fields $\s_n$ which effectively define the $n$-twisted sectors; for $n$ odd, $\s_n$ is the lowest weight field of the sector, but for even $n$ the lowest weight fermionic operators are the $R^\pm_n$ and $R^{0\pm}_n$.

The ring of NS chiral operators, which have R-charges equal to their conformal weights, can be built  with an even twist by applying fractional R-current modes on these lowest weight states \cite{Lunin:2001pw}.
Applying the modes to the R-charged fields $R^+_n$ and $R^-_n$, one obtains the chiral operators $O^{(0,0)}_n$ and $O^{(2,2)}_n$ with the highest and the lowest conformal weights, respectively, i.e.~$h = \frac{n \pm 1}{2} = j^3$, where $j^3$ is the holomorphic R-charge. Meanwhile, starting with the R-neutral doublet $R^{0\pm}_n$  with $n$ even, one can create middle-cohomology chiral fields $O^{(1,1)\pm}_n$, both with $h = \frac{n}{2} = j^3$, and distinguished by the internal SU(2)$_2$ charge.
Chiral operators are protected against the deformation (\ref{def-cft}) as they saturate a BPS bound, so their dimensions and charges do not change as one moves in moduli space. In fact, the single-cycle chiral operators correspond to single-particle excitations of the supergravity solutions, and the matching between three-point functions of chiral fields in the free orbifold CFT and the corresponding correlators computed from the asymptotically AdS solutions, even as each of the two descriptions hold in separate points of moduli space, is a remarkable success of the holographic correspondence
\cite{Lunin:2001jy,Kanitscheider:2006zf,Kanitscheider:2007wq,Taylor:2007hs,Baggio:2012rr,deBoer:2008ss}.

The chiral operators can also be related to the Ramond fields by spectral flow of the $n$-twisted algebra, with central charge $c = 6n$;
for example,  $O^{(1,1)\pm}_n$ flows to $R^{0\pm}_n$.
At first sight, this could suggest that $R^{0\pm}_n$ is protected but, as discussed in \cite{Lima:2020kek}, $\Oint$ changes the length of the $n$-cycle, so one cannot perform the specific spectral flow with $c = 6n$ in the deformed theory.
The Ramond ground states of the full orbifold CFT, with $h = \frac14 N$ (corresponding to $c = 6N$), are products of single-cycle R-neutral and/or R-charged Ramond fields $R^{(i)}_{n_i}$, with the cycles forming a conjugacy class of the full $S_N$, i.e.~$\prod_i (R^{(i)}_{n_i})^{k_i}$
with $\sum_i k_i n_i = N$.
\emph{These} products are protected, as shown in \cite{Lima:2021wrz}.
They are the CFT translations of two-charge geometries of the D1-D5 system%
\footnote{%
There are similar constructions for global AdS$_3 \times S^3$ geometries made by products of NS chiral fields
\cite{Lunin:2002bj,Lunin:2002fw}.
}
under a holographic dictionary
\cite{Lunin:2001jy,Kanitscheider:2006zf,Kanitscheider:2007wq,Skenderis:2008qn} but, while all of the ground states contribute to the microscopic entropy of the system, only those with a definite (and large) R-charge have been associated with smooth, horizonless, non-singular microstate geometries
in the low energy approximation of supergravity
\cite{Kanitscheider:2006zf,Skenderis:2008qn}.
For example, the simplest of such geometries \cite{Lunin:2001jy} is dual to
$( R^-_n)^{N/n}$.
For the total R-charge to be large, there must not be too many R-neutral fields entering the superposition; in particular, superpositions made with only R-neutral fields, say
$(R^{0\pm}_{n})^{N/n}$,
have exactly zero R-charge. This means that the external SO(4) symmetry of $S^3$ in the asymptotically AdS$_3 \times S^3$ background is unbroken and, as a consequence, the corresponding geometries, which have only internal excitations, are indistinguishable in the supergravity approximation \cite{Skenderis:2008qn}.
That is, perhaps, one of the reasons why the R-neutral fields are less studied than the R-charged ones. This work gives a modest contribution in this regard, as we examine some properties both of $R^{0\pm}_n$ and of the chiral fields $O^{(1,1)\pm}_n$.

The strategy we use to study the effect of $\Oint$ on the R-neutral Ramond fields is to compute
\be
\Big\langle R^{0+}_{[n]}(\infty, \bar \infty) \Oint(1,\bar 1) \Oint(u,\bar u) R^{0-}_{[n]}(0,\bar 0) \Big\rangle . \label{4ptIntro}
\ee
The $[n]$ indicates the $S_N$-invariant combination of twists.
$N$-dependent factors coming from combinatoric properties of the permutations involved in (\ref{4ptIntro}) can be organized in an $1/N$-expansion similar in spirit to the 't Hooft expansion, with the large-$N$ overall scaling behavior of the function being $N^{-{\bf g} - 1}$, where $\bf g$ is the genus of the covering surface used to compute the correlator, related to the permutations by the Riemann-Hurwitz formula \cite{Lunin:2000yv,Pakman:2009zz}. 
From the brane system perspective, it is natural to take $N$ large, and here we restrict ourselves to the leading-order contribution, that is, to genus-zero covering surfaces.
The four-point function is a dynamical object, not fixed by conformal symmetry, and it can be used to extract relevant information about the interaction of the fields.
Taking the appropriate limits, one can find OPEs between operators, along with the corresponding structure constants. Thus, expanding the function for $u \to 0$ gives the OPE
\be \label{OPEintro}
[ \Oint ] \times [ R^{0\pm}_{[n]} ]
\ee
which contains the operators resulting from acting with $\Oint$ on $R^{0\pm}_{[n]}$.
Similar computations of four-point functions with two $\Oint$ insertions have been examined in \cite{Burrington:2012yq}, with some selected non-twisted operators entering in the place of the Ramond fields in (\ref{4ptIntro}), while
in \cite{deBeer:2019ioe}, four-point functions with only chiral operators but the same twist structure as (\ref{4ptIntro}) were computed.
As discussed in those references, the use of four-point functions as the starting point to obtain three-point functions and structure constants has the (very) pragmatic effect of selecting only the non-vanishing three-point functions that appear in the OPE. Finding three-point functions (i.e.~structure constants) with one $\Oint$ insertion gives information about mixing at linear order.
Furthermore, working in $\la$-perturbation theory, integrating the function (\ref{4ptIntro}) over the position of the interactions gives us the anomalous dimension of $R^{0\pm}_n$, to second-order in $\la$. While the first-order correction vanishes, we show, following the regularization developed in \cite{Lima:2020boh,Lima:2020kek}, that the second-order correction does not, so $R^{0\pm}_n$ is lifted at order $\la^2$, and at the leading order in the large-$N$ expansion. The lifting occurs for $n < N$; the dimension of R-neutral Ramond fields with maximal twist $n=N$ is protected, at least at leading order in $1/N$.
In \cite{Pakman:2009mi}, a function similar to (\ref{4ptIntro}), but with the lowest-weight $n$-twisted chiral primary $O^{(0,0)}_n$ was computed and integrated to show that the second-order correction to the dimension $h = \frac{n-1}{2}$ vanishes, as expected for a protected chiral field.
Here we compute the corresponding function for the chiral operators $O^{(1,1)\pm}_n$, with $h = \frac{n}{2}$, related to the R-neutral fields $R^{0\pm}_n$ by spectral flow, and we show that its integral also vanishes.

Although the present results --- the OPE (\ref{OPEintro}) and the renormalization of the dimension of $R^{0\pm}_n$ ---
follow the same qualitative pattern as the ones in \cite{Lima:2020boh,Lima:2020kek}, there are relevant technical differences in both cases.
Some fortunate idiosyncratic cancellations occur in the derivation of the four-point function in the R-charged case of Refs.\cite{Lima:2020boh,Lima:2020kek}, which make them quite simpler to obtain than the one derived here.
In contrast, the present computation of (\ref{4ptIntro}) follows a rather general pattern that can be followed directly in the derivation of similar functions involving other operators instead of the Ramond fields. Furthermore, in contrast to \cite{Lima:2020boh,Lima:2020kek}, here we strive for a little more clarity by computing the function in two different ways.

As mentioned above, the computation of twisted correlation functions such as (\ref{4ptIntro}) is, by itself, a non-trivial work, because one must take the complicated monodromies into account. There are two main methods for doing this. One is the stress-tensor method, first introduced in \cite{Dixon:1986qv} for $Z_n$ orbifolds, and later used in \cite{Arutyunov:1997gi,Arutyunov:1997gt,Pakman:2009zz,Pakman:2009ab,Pakman:2009mi} for $S_N$ orbifolds.
The other way is the Lunin-Mathur (LM) technique \cite{Lunin:2000yv,Lunin:2001pw}, which consists of evaluating the Liouville contribution from the twists  to the path integral, with the aid of the appropriate ramified covering surface with genus $\bf g$.
While the LM technique has been widely used \cite{Burrington:2012yq,Burrington:2012yn,Burrington:2015mfa,Burrington:2017jhh,Tormo:2018fnt,Burrington:2018upk,deBeer:2019ioe}, our recent results  for  R-charged twisted Ramond fields \cite{Lima:2020boh,Lima:2020kek} were obtained with the stress-tensor method.
In the present paper, we will compute (\ref{4ptIntro}) using both approaches, which gives an interesting opportunity for seeing how they are complementary.
The most complicated aspect of the LM technique is that the computation of the Liouville factor involves a regularization procedure of ``closing holes'' around the branching points which must be carefully done so that the final, physical amplitudes are finite and well-defined. But, once the Liouville factor is computed, all that remains is to compute a simple correlation of free fields on the covering surface. The stress-tensor method, on the other hand, does not require any regularization at all; however, instead of computing the desired correlation function directly, one first determines a first-order differential equation which has to be integrated. When applied to the bare twist fields correlation function, the stress-tensor method becomes quite trivial \cite{Lima:2020kek}, and here, when we compute the Liouville factor for the same covering map (with a specific parameterization and for ${\bf g} = 0$, i.e.~in the large-$N$ limit), we can check that both results agree. In short, it is interesting to see how the stress-tensor method can be used to go around the complicated regularization of the Liouville factor, while the ``rest'' of the LM technique can be used to give a direct formula for the full correlation, bypassing the need to solve a differential equation.

\bigskip

The structure of the paper is as follows. 
In Sect.\ref{orbi} we review the most relevant facts about the D1-D5 SCFT, both at the free-orbifold point and at its deformation, and fix our notations.
In Sect.\ref{SectTwoPointFuncts} we compute the four-point function (\ref{4ptIntro}) in detail, first via the Lunin-Mathur technique, and then with the stress-tensor method.
In Sect.\ref{OPEsingle} we use the coincidence limits of the function to extract OPEs, including (\ref{OPEintro}). 
We also present the similar four-point function with middle-cohomology NS chirals $O^{(1,1)\pm}_{[n]}$ in place of the Ramond fields, extract the OPE analogous to (\ref{OPEintro}), and compare the two results.
In Sect.\ref{RenormalizationNeutral}, we integrate (\ref{4ptIntro}) using the convenient regularization procedure, and obtain the anomalous dimensions of the renormalized Ramond fields, at order $\la^2$ and for large $N$; we do the same with the four-point function involving $O^{(1,1)\pm}_{[n]}$, and verify that this field is protected, as it should be.
We conclude in Sect.\ref{SectConcl}. 
App.\ref{SectLMRcharged} contains a brief computation of the correlator of the R-charged Ramond fields using the Lunin-Mathur technique, using some results derived in Sect.\ref{SectTwoPointFuncts}.

\section{The D1-D5 SCFT} \label{orbi}

The `free point' in the  moduli space of D1-D5 SCFT  is a symmetric product orbifold, made by $N$ copies of the $\cal N = (4,4)$ super-conformal field theory on the torus $T^4$, identified under the action of the symmetric group $S_N$, resulting in the orbifold  target space $(T^4)^N/S_N$.
Each copy contains four free scalar bosons $X^i_I(z,\bar z)$, and four free fermions $\psi^{i}_I(z)$, where $i=1, \dots ,4$ labels the fields and  $I=1,\dots,N$ the copies; the total central charge with the $4N$ bosons and $4N$ fermions is $c_{orb}=6N$.  It is convenient to pair the real bosons $X^i_I$ into complex bosons $X^a_I$ and $X^{a\dagger}_I$,  and the Majorana fermions into complex fermions $\psi^a_I(z)$, with $a = 1,2$.
In what follows, we always work with the $X^a$ and $X^{a\dagger}$, and we bosonize the complex fermions with $2N$ new free bosons $\phi^a_I(z)$, 
\be
\psi_I^a(z) = e^{i\phi_I^a}(z) ,  \qquad \psi_I^{a\dagger} (z) =e^{-i\phi_I^a}(z) .
\ee

The holomorphic%
\footnote{We work with $z,\bar z$, defined on the complex plane.}
 $\cal N=4$ super-conformal symmetry  is generated by the stress-energy tensor $T(z)$, the SU(2) R-currents $J^r(z)$, $r=1,2,3$, and the super-currents $G^a(z)$, $\hat G^a(z)$, which can be expressed in terms of the free fields as
\bsub
\begin{align}
T(z) &= - \tfrac12 \lim_{w\rightarrow z}
\textstyle\sum_{I=1}^N
\left( \pa X^a_I(z) \pa X^{a\dagger}_I(w)
+ \pa \phi^a_I (z) \pa \phi^a_I(w)
+\frac{6}{ (z-w)^2}\right)
\label{stres}
\\
J^3(z) &= \tfrac{i}{2} \textstyle\sum_{I=1}^N (\p\phi^1_I+\p\phi^2_I)
\label{j3}
\\
{ G}^1(z)
&=
i \sqrt2  
\textstyle\sum_{I=1}^N
\big( \psi^1_I \pa  X_I^{1\dagger}
+ \psi^{2}_I \pa  X_I^{2\dagger}
\big) ,
\\
{ G}^2 (z) &=
\sqrt2  
\textstyle\sum_{I=1}^N
\big(
\psi^{1\dagger}_I \pa  X_I^{2\dagger}
- \psi^{2\dagger}_I \pa  X_I^{1\dagger}  
\big)
\end{align}\esub
along with $\hat G^a (z) = - G^{a\dagger}(z)$.
The  eigenvalues of the zero-mode of the current $J^3(z)$ define the R-charge $j^3$.
The anti-chiral currents, $\tilde T(\bar z)$, $\tilde J^3(\bar z)$, $\tilde G^a(\bar z)$, $\tilde{\hat G}^a(\bar z)$, have analogous forms in terms of the right-moving fields.

For each CFT copy, the four bosons $X^i_I(z,\bar z)$ are coordinates on the torus $T^4$ and, although the periodic identifications break the rotational symmetry of four-dimensional Euclidean space, it is convenient to use this broken ``internal'' symmetry group $\rm{SO}(4)_I = \rm{SU}(2)_1 \times \rm{SU}(2)_2$. 
The complex fermions transform as doublets of the SU(2)$_2$ factor, which is an automorphism of the superconformal algebra. More precisely, holomorphic fermions transform as doublets of $\rm{SU}(2)_L \times \rm{SU}(2)_2$, and anti-holomorphic ones as doublets of  $\rm{SU}(2)_R \times \rm{SU}(2)_2$, where SU(2)$_{L,R}$ are the R-symmetry groups. 
Note that the same SU(2)$_2$ acts on both sectors, so the corresponding fermionic charges are the eigenvalues of the ``total'' conserved current 
$\frak J(z) + \tilde{\frak J}(\bar z)$.
After bosonization, the holomorphic contribution is written as
\be\label{j3til}
\frak J^3(z)= \tfrac{i}{2} \textstyle\sum_{I=1}^N (\p\phi^1_I - \p\phi^2_I) (z)
\ee
and the anti-holomorphic one as
$\tilde{\frak J}^3(z)= \tfrac{i}{2} \textstyle\sum_{I=1}^N (\p\tilde\phi^1_I - \p\tilde\phi^2_I) (\bar z)$.
We will be somewhat lax with our nomenclature, usually considering just $\frak J(z)$ explicitly, as the treatment of $\tilde{\frak J}^3(\bar z)$ is analogous.
Note that the SU(2) currents all have corresponding raising and lowering components as well, 
$J^\pm(z) = e^{\pm i \sum_I (\phi^1_I + \phi^2_I)}$, 
${\frak J}^\pm(z)= e^{\pm i \sum_I (\phi^1_I - \phi^2_I)}$, etc.,
creating the doublets.

The ground states of the orbifold SCFT are organized in twisted sectors with all the allowed  $S_N$-boundary conditions, which can be realized by the insertion of  `twist fields' $\sigma_g(z,\bar z)$ for each $g \in S_N$, such that,
e.g.,~
$$
X^i_I(e^{2\pi i}z,e^{-2\pi i}\bar z) \s_g(z,\bar z) =X^i_{g(I)}(z,\bar z) \s_g(z,\bar z).
$$
Conjugacy classes of $S_N$  are in one-to-one correspondence with the subgroups of cyclic permutations $Z_n$, $n=1, \dots ,N$.
In this paper, we will be interested in the simplest, single-cycle permutations  corresponding to cycles $(n)$ of length $n$.
To obtain an $S_N$-invariant operator belonging to the conjugacy class $[n]$ of length-$n$ cycles, we sum over the orbits of $(n)$, and denote the resulting operator as $\sigma_{[n]}\sim \sum_{h\in S_N}\sigma_{h^{-1}(1,\dots,n)h}$. The conformal weights of any single-cycle field $\s_n(z,\bar z)$, hence also of $\s_{[n]}(z,\bar z)$, are given by \cite{Dixon:1986qv}
\be\label{twistdim}
h^{\sigma}_n= \frac{1}{ 4} \left( n- \frac{1}{ n} \right) = \tilde h^\s_n .
\ee

\subsection{Twisted fermions}	\label{SectTwistedFerm}

When fermions are involved, we must consider their periodicity, along with the twisted boundary conditions introduced by the orbifold, and then the notion of a strictly periodic or anti-periodic fermion loses its meaning somewhat  \cite{Avery:2009xr,Lunin:2001pw}. 
Let us elaborate on this point, since is important for motivating our main computation. We follow closely a discussion made in Ref.\cite{Avery:2009xr}.
In the seed CFT, we can parameterize periodicity by a phase $\tau_{a}$ such that
\be
\psi^a(e^{2\pi i}z) = e^{i\pi \tau_a} \psi^a(z) 
\quad
	\begin{cases}
	\tau_a = 1 \qquad& \text{Ramond}
	\\
	\tau_a = 0 \qquad& \text{Neveu-Schwarz}
	\end{cases}
\ee
We will omit the label $a$ in $\tau$ from now on. In the $n$-twisted sector, going around the twist has the effect of not only changing a phase, but also swapping the field by a different one
\be
\begin{split}
\psi_I^a(e^{2\pi i}z) \s_{(1,\cdots,n)}(0) 
	&= e^{i\pi \tau} \psi_{I+1}^a(z) \s_{(1,\cdots,n)}(0) ,
	\qquad
	1 \leq I \leq n-1
\\
\psi_n^a(e^{2\pi i}z) \s_{(1,\cdots,n)}(0) 
	&= e^{i\pi \tau} \psi_{1}^a(z) \s_{(1,\cdots,n)}(0) ,
\end{split}
\ee
so we see that the notion of a periodic or anti-periodic fermion becomes ill-defined.
We can return to the same field if we repeat this operation $n$ times, resulting in
\be	\label{psirndn}
\psi_I^a(e^{2 n \pi i}z) \s_{(1,\cdots,n)}(0) 
	= e^{in \tau \pi} \ \psi_I^a(z) \s_{(1,\cdots,n)}(0)  .
\ee
This is the most general way of defining the boundary conditions for the fermions.

A powerful way of disentangling the boundary conditions of the $n$-twisted sector is to map the ``base sphere'' to a covering surface $\Sigma$ with a branching point of order $n$ at the pre-image $t$ of the insertion point $z$ of each twist $\s_n (z)$ \cite{Lunin:2000yv,Lunin:2001pw}.
The ramified structure of the map 
\be	\label{zt0}
z(t) \approx b_0 t^n
\ee
in the vicinity of $z = 0$ implements the twisted boundary conditions in such a way that, on $\Sigma$, there is only one copy of the basic fields, i.e.~a CFT with $c = 6$, and the twist $\s_n |_{z=0}$ is lifted to the identity ${\mathds 1}|_{t=0}$. Since fermions have weight $h = \frac12$, they lift to the covering surface as%
	\footnote{%
	Apart from factors of $b_0$, see the discussion around Eq.(\ref{rim}) later. These factors do not matter here since they cancel in (\ref{psirsc}).
	}
$
\psi_I^a(z) \mapsfrom (dz/dt)^{- \frac12} \psi^a(t)
$.
Note the the copy index disappears. 
Hence lifting each side of Eq.(\ref{psirndn}) with (\ref{zt0}) we find
\be	\label{psirsc}
\psi^a(e^{2 \pi i} t) \; {\mathds 1}(0) 
	= e^{in \tau \pi + i (n-1)\pi} 
		\, \psi^a(t) \; {\mathds 1}(0)  .
\ee
We now do have a definite periodicity for the single fermion living on $\Sigma$, with a phase corrected by the factors of $(dz/dt)^{-\frac12}$, 
\begin{align}
e^{i [ n(\tau +1) - 1] \pi}
	= 
	\begin{cases}
	-1 \quad&		\text{for $\tau = 1$ and $\forall \ n$}
\\
	+1 \quad&		\text{for $\tau = 0$ and odd $n$}
\\
	-1 \quad&		\text{for $\tau = 0$ and even $n$}
	\end{cases}
\end{align}
The first case, with $\tau = 1$, corresponded to the Ramond sector in the seed CFT, and we see that it still corresponds to the Ramond sector of the covering CFT. The two last cases, with $\tau = 0$, corresponded to the NS case of the seed CFT, but we see that it only corresponds to the NS sector of the covering CFT if $n$ is odd; for even $n$, the periodic boundary conditions in the seed CFT are ``mapped'' to anti-periodic conditions in the covering CFT.%
	\footnote{%
	We put quotation marks in ``mapped'' because the concept of a seed CFT is only auxiliary.}

In any case, the anti-periodic boundary conditions on the covering surface can be implemented by the insertion of `spin fields', which create the degenerate set of Ramond ground states of the $c = 6$ SCFT. There are four holomorphic spin fields, $R^{\pm}(t)$ and $R^{0\pm}(t)$, all with the  conformal weights 
$(h^\R, \tilde h^\R) = (\frac{c}{24} , 0) = (\frac14 , 0)$; as well as four anti-holomorphic ones, $\tilde R^{\pm}(\bar t)$, $\tilde R^{0\pm}(\bar t)$, with $(h^\R , \tilde h^\R) = (0, \frac14 )$.
The (holomorphic) spin fields are distinguished by their charges under the R-current $J^3(t)$ and the ``internal'' current ${\frak J}^3(t)$, defined as in (\ref{j3}) and (\ref{j3til}) without the sums over copies, as there is only one copy on the covering. 
Inserting a spin field on the covering surface is tantamount to inserting a corresponding operator, which we call a `Ramond field', on the base.
The four holomorphic Ramond fields in the $n$-twisted sector are given by
\begin{flalign}\label{micro1}
&& &R_n^{\pm}(z)= 
 	e^{\pm \frac{i}{2n}\sum_{I=1}^n \left[ \phi_{I}^1(z)+\phi_{I}^2(z) \right]}
	\sigma_{(1,\cdots,n)}(z) &&
\\
\text{with}
&&  &h^R_{n} = \tfrac{1}{4} n = \tilde h^R_n , 
\quad j^3 =\pm \tfrac{1}{2} ,
\quad \frak j^3 = 0 ,
\label{microdim}
\end{flalign}
which form a doublet of the SU(2) R-symmetry generated by the current $J^i(z)$, and  
\begin{flalign}\label{micro0}
&& &R_n^{0\pm}(z)= 
 	e^{\pm \frac{i}{2n}\sum_{I=1}^n \left[ \phi_{I}^1(z)-\phi_{I}^2(z) \right]}
	\sigma_{(1,\cdots,n)}(z) &&
\\
\text{with}
&&  &h^R_{n} = \tfrac{1}{4} n = \tilde h^R_n , 
\quad j^3 = 0 ,
\quad \frak j^3 = \pm \tfrac{1}{2} ,
\label{dimmicro0}
\end{flalign}
which R-neutral, and distinguished by the internal SU(2) symmetry generated by $\frak J^i(z)$. 
We can construct $S_N$-invariant operators by summing over the group orbit of the cycle. Explicitly, for the R-neutral doublet, we have
\be\label{microc}
R^{0\pm}_{[n]}(z) \equiv 
	\frac{1}{\scr S_n(N)}
	\sum_{h \in S_N}
	\exp \left( \pm \frac{i}{2n} \sum_{I=1}^n \big[ \phi^1_{h(I)}(z) - \phi^2_{ h(I)}(z)  \big] \right) \s_{h^{-1} (1 \cdots n) h}(z)
\ee
where the factor $\scr S_n(N)$ is such that the two-point function is normalized.
Of course, the $S_N$-invariant fields have the same quantum numbers as the corresponding non-$S_N$-invariant ones.
We will use basically the same notation for left-moving  as well as for left-right moving fields, usually distinguishing both by the argument e.g.
\be
R^{0\pm}_n(z,\bar z) 
	=
 	e^{\pm \frac{i}{2n}\sum_{I=1}^n \left[ \phi_{I}^1(z)-\phi_{I}^2(z) \right]}
	e^{\pm \frac{i}{2n}\sum_{I=1}^n \left[ \tilde\phi_{I}^1(\bar z)- \tilde\phi_{I}^2(\bar z) \right]}\sigma_{(1,\cdots,n)}(z,\bar z) .
\ee

In the twisted sectors there is a ring of NS chiral operators with $(h,\tilde h) = (j^3 , \tilde \jmath^3)$. The lowest- and highest-weight chirals  for twist $n$ are $O^{(0,0)}_n$ and $O^{(2,2)}_n$, with $h = \frac{n\mp1}{2} = j^3$.
They are related to the R-charged Ramond fields $R^-_n$ and $R^+_n$, respectively, by spectral flow of the $n$-twisted cyclic orbifold CFT with $c = 6n$. The same spectral flow applied to the R-neutral fields $R^{0\pm}_n$ results in the `middle-cohomology' chirals $O^{(1,1)\pm}_n$, both of which have $h = \frac{n}{2} = j^3$, and are distinguished by their charge $\frak j^3 = \pm \frac12$.
One can also obtain the chiral operators by starting with the lowest-weight fermionic field with the lowest R-charge in the $n$-twisted sector, and filling a Fermi sea by applying  fractional modes $\Psi_{- \frac{k}{2n}}$ of the 
twist-invariant fermion 
$\Psi^a = \sum_{I = 1}^n \psi^a_I$,
thus raising the charges and the weight until one reaches $h = j^3$ \cite{Lunin:2001pw}. 
A convenient representation of the chiral operators can be given in the bosonized language, see e.g.~\cite{Pakman:2009ab,Pakman:2009mi}; for example, the middle-cohomology operators are
\be	\label{O11pm}
\begin{split}
& O^{(1,1)\pm}_n
	= 
 	e^{ \frac{i}{2}\sum_{I=1}^n 
			\left[ 
				\frac{n \pm 1}{n} \phi_{I}^1(z) 
				+ 
				\frac{n \mp 1}{n} \phi_{I}^2(z) 
			\right]}
	e^{ \frac{i}{2}\sum_{I=1}^n 
			\left[
				\frac{n \pm 1}{n} \tilde\phi_{I}^1(\bar z) 
				+
				\frac{n \mp 1}{n} \tilde\phi_{I}^2(\bar z) 
			\right]}
	\sigma_{(1,\cdots,n)}(z,\bar z) 
\\
& (h , \tilde h) = (\tfrac12 n , \tfrac12 n) = (j^3 , \tilde \jmath^3) \ ,
\qquad
(\frak j^3 , \tilde{\frak j}^3) = ( \pm \tfrac12 , \pm \tfrac12) ,
\end{split}
\ee
with a corresponding $S_N$-invariant field given by a sum over orbits as in Eq.(\ref{microc}).
Note that the middle cohomology chiral operators associated with the R-neutral Ramond fields are characteristic of the internal $T^4$ manifold; the other possible compactification of the D1-D5 system, with internal K3, has a different set of such operators. Meanwhile, the chiral fields associated with the R-charged Ramond fields are universal in this sense.

Let us recapitulate. 
	If $n$ is odd, we can have periodic or anti-periodic boundary conditions on the covering surface;
	if $n$ is even, there is no well-defined notion of a periodic fermion in the $n$-twisted sector. 
	Hence we can either choose ($n$ odd) or be forced ($n$ even) to insert Ramond fields, which are therefore fundamental to the definition of the ground states. 
Given the lowest-weight, lowest R-charged fermionic operator, which for even $n$ is $R^-_n$, one can obtain the set of chiral operators by applying fractional modes of fermions; in this way (and using the relations between fermions and spin fields), the middle-cohomology operators $O^{(1,1)\pm}_n$ can be obtained from $R^{0\pm}_n$.

We have recently shown in \cite{Lima:2020boh} that the bare dimension 
$\Delta^R_n = h^R_n + \tilde h^R_n$ 
of the R-symmetry doublet (\ref{micro1}) is renormalized when the free orbifold theory is deformed.
One of the goals of the present paper is to show that the internal SU(2) doublet (\ref{microc}) is renormalized as well.

\subsection{Deformation}

In the deformed theory, the  scalar modulus  interaction operator has to be  marginal, i.e.~of conformal dimension $\Delta=2$, to preserve the $\cal N= (4,4)$ supersymmetry and to be invariant under the SU(2) symmetries.  Its explicit form is known to be 
\be\label{interaction}
\Oint (z,\bar z) = 
	 \frac{i}{2} \left(  
	 G^1_{- \frac12} \tilde G^2_{- \frac12}
	 - G^2_{-\frac12} 
	 \tilde{ G}^1_{-\frac12}
	 \right) O^{(0,0)}_{[2]} (z,\bar z)+ c.c.
\ee
The NS chiral operators
$O^{(0,0)}_{[n]}(z,\bar z)$, which have conformal weight and R-charge
$h = \frac{n-1}{2} = j^3$, are the lowest-weight operators in the chiral ring for twist $n$. In (\ref{interaction}), we have a descendant of $O^{(0,0)}_{[2]}$, with twist $n=2$, whose total dimension after applying the supercharges is $\Delta^{\rm{int}} =  2$. 

We are interested in the description of the  large-$N$ properties of the twisted Ramond fields in the deformed orbifold SCFT (\ref{def-cft}), up to second order in the perturbation theory for the deformation parameter $\la$ and  
in particular, in the calculation of the corrections to their bare conformal dimension. 
The first-order correction to $\Delta^R_n$ vanishes, because it is given by the structure constant in the three-point function $\langle R^{0-}_{[n]} \Oint R^{0+}_{[n]} \rangle = 0$.
The second-order correction  is given by the integral
\be	\label{Int}
D =
\frac{\la^2}{ 2} \int\! d^2z_2 \int \! d^2z_3 \,  
	\Big\langle 
		R^{0-}_{[n]} (z_1,\bar z_1)
		\Oint (z_2, \bar z_2) 
		\Oint (z_3, \bar z_3) 
		R^{0+}_{[n]} (z_4, \bar z_4) 
	\Big\rangle.
\ee
Using conformal invariance,  one can bring the four-point function under the integral to the form
\be\label{confor}
	\Big\langle 
		R^{0-}_{[n]} (z_1,\bar z_1)
		\Oint (z_2, \bar z_2) 
		\Oint (z_3, \bar z_3) 
		R^{0+}_{[n]} (z_4, \bar z_4) 
	\Big\rangle
	=
	|z_{13}z_{24}|^{-4}|z_{14}|^{-n+4}G_0(u,\bar u)
\ee
where $u= ( z_{12}z_{34}) / (z_{13}z_{24})$ and%
	\footnote{%
	The index 0 is to emphasize that this function contains R-neutral Ramond fields.
	}
\be\label{gu}
G_0(u,\bar u)
	\equiv
	\Big\langle 
		R^{0-}_{[n]} (\infty,\bar \infty)
		\Oint (1, \bar 1) 
		\Oint (u, \bar u) 
		R^{0+}_{[n]} (0,\bar 0) 
	\Big\rangle .
\ee
After a change of integration variables (\ref{Int}) becomes
\be
\frac{1}{ |z_{14}|^{4h}} 
\int \! d^2 w \frac{1}{ |w|^2||1+w|^2}
	\int \! d^2 u \, G_0(u,\bar u) .
\ee
The remaining integral over  $w={z_{13} / z_{14}}$ is divergent, 
and must be regularized by a UV cutoff $\La$, 
$\int \! d^2 w \frac{1}{ |w|^2||1+w|^2}=2\pi \log\Lambda$,
resulting in
\be
D = \lambda^2\pi \frac{\log\Lambda}{ |z_{14}|^n} J(n),\label{correct}
\ee
where 
\be
J(n) \equiv \int \! d^2u \, G_0(u,\bar u).
\ee
The logarithmic dependence on the cutoff $\La$ is the hallmark of the change in the conformal dimension of the deformed Ramond fileds  in the renormalized two-point function \cite{Lima:2020kek}
\bsub	\label{anomdim}
\begin{align}
& \Big\langle R^{0-}_{[n]}(z_1, \bar z_1) R^{0+}_{[n]}(z_2 , \bar z_2) \Big\rangle_\la
	= |z_{12}|^{-2 \Delta_\la}
,
\\
& \Delta_\la(n) = \Delta^R_n - \tfrac{\pi}{2} \la^2 J(n) + \rm{O}(\la^3) ,	\end{align}
\esub
where $\Delta^R_n = h^R_n + \tilde h^R_n = \frac12 n$ is the bare dimension of $R^{0\pm}_{[n]}$. 
The proper  regularization of $J(n)$ and its explicit evaluation is one of the problems addressed in the present paper.

\section{Computation of the four-point function}	\label{SectTwoPointFuncts}

In order to derive the second-order correction  (\ref{anomdim}) to the conformal dimensions of the R-neutral  twisted Ramond fields,  we have to first  calculate the four-point function $G_0(u,\bar u)$ given in (\ref{gu}). 
It is clear that this function should be multi-valued due to the orbifold boundary conditions. The most convenient way of computing it is to use the covering surface \cite{Lunin:2000yv,Lunin:2001pw} described in \S\ref{SectTwistedFerm}. For the correlator in question, the covering map from the genus-zero covering surface, with coordinates $(t,\bar t)$, to the base sphere, with coordinates $(z,\bar z)$, can be parametrized as in Refs.\cite{Arutyunov:1997gt,Pakman:2009mi,Lima:2020kek}
\be\label{cover}
z(t)=\left( \frac{t}{ t_1}\right)^n \left( \frac{t-t_0}{t_1-t_0} \right) \left( \frac{t_1-t_\infty}{ t-t_\infty} \right).
\ee
By construction this map has correct monodromies around the images $z = \{0,\infty\}$ of the covering points $t = \{ 0,t_0 ; t_\infty,\infty\}$, where $n$-twists are inserted. To ensure the correct branching around the insertions of twists $n=2$ fields at the points $z(t_1) =1$ and $z( x ) \equiv u$,
we impose the conditions $z - z_* \sim (t-t_*)^2$ for $t = t_1$ and $t = x$.
This fixes the coefficients in (\ref{cover}) as functions of $x$, which can be put in the form
\begin{align}	\label{tis}
t_0=x-1,
\quad
t_{\infty}=x- \frac{x}{x+n},
\quad
t_1={1-n\over n}+x- \frac{(n+1)x}{n(x+n)} 
\end{align}
With this choice, we  get the final form of the $u = z(x)$  parametrization
\be\label{ux}
u(x)= \frac{x^{n-1}(x+n)^{n+1}}{ (x-1)^{n+1}(x+n-1)^{n-1}}.
\ee
The map (\ref{cover}) assures that  the covering surface has the topology of a sphere. 
In general, the twisted four-point function will have contributions from coverings with higher genera, but for large $N$ the genus-zero contribution is the leading one \cite{Lunin:2000yv,Pakman:2009zz}, and it can be shown that $G_0(u,\bar u)$ vanishes at this leading order when $n = N$.

\subsection{The Lunin-Mathur technique}		\label{SectLMtech}

Perhaps the most standard way of computing $G_0(u,\bar u)$ is to use the Lunin-Mathur (LM) technique \cite{Lunin:2000yv,Lunin:2001pw}.
It uses the fact that functional integrals $Z_{\rm{base}}$ and $Z_{\rm{cover}}$ for correlation functions on the base and on the covering surfaces are related by a Liouville factor, $Z_{\rm{base}} = e^{S_L}  Z_{\rm{cover}}$, where $S_L$ is the Polyakov-Liouville action for the Weyl transformation of the metrics, $ds^2_{\rm{base}} = e^\phi d s^2_{\rm{cover}}$.
The path integral computation makes it evident that the correlation function (\ref{gu}) factorizes as
\be
G_0 = G_\s \times G_B \times G_F ,	\label{GsBF}
\ee
(or possibly a sum of terms like these)
where $G_B$ is the path integral for the bosons, $G_F$ is the correlator for the fermions, and $G_\s$ the correlation for the twists. 
The latter is given by the Liouville factor, and the bosonic and fermionic functions are the non-twisted correlation functions at the covering surface.

The twist factor 
\be	\label{gsig}
G_\s = \big\langle \s_{[n]}(\infty, \bar \infty) \s_{[2]}(1, \bar 1) \s_{[2]}(u,\bar u) \s_{[n]}(0,\bar 0) \big\rangle = e^{ S_L }
\ee
is fixed by the choice of the covering map, and is universal for all correlation functions with the same twist structure. The function for the specific twists (\ref{gsig}) was given by LM in \cite{Lunin:2000yv}; see also \cite{deBeer:2019ioe} for a function with dressed twisted operators with the same twist structure, and \cite{Avery:2010qw} for a general analysis.
It is nevertheless instructive to show here how to compute $G_\s$, since the parameterization (\ref{tis})  of the covering map is different from the one in \cite{Lunin:2000yv}, resulting in a different form for $G_\s$. In the vicinity of a point $z_*$ with a twist $\s_{n_*}$, the covering map has the structure  
\be\label{coefb}
z(t) = z_* + b_* ( t - t_*)^{n_*} + \cdots
\ee
and the parameters $b_*$ and $n_*$ fix the Liouville action as%
	\footnote{%
	See Eq.(D.63) of \cite{Avery:2010qw}.
	}
\begin{align}
\begin{split}
S_L = 
	- \frac{c_{\rm{cover}}}{12} 
	&\Bigg[ 
	\sum_* \frac{n_*-1}{n_*} \log | b_*| 
	+ 
	\frac{n_{t_\infty} +1}{n_{t_\infty}} \log |b_{t_\infty}|
	- 
	\frac{n_\infty - 1}{n_\infty} \log | b_{\infty}|
\\
&
	+
	\sum_* (n_* -1) \log n_* 
	-
	(n_{t_\infty} +1) \log n_{t_\infty}
	-
	(n_\infty + 3) \log n_\infty
\\
&
	+ \text{Regulation terms}	
	\Bigg]
\end{split}
\end{align}
The `regulation terms' are singular terms depending on the the log of the small regulating parameters used to cut discs around the singular ramification/branching points. When proper, careful account is taken of these terms, one can define a correctly normalized twist operator such that they vanish in a given correlation function \cite{Lunin:2000yv}.
Note that there is a distinction between the contribution of the region $|t| = \infty$, where we define $z \approx b_\infty t^{n_\infty}$, and the finite points%
	\footnote{%
	Here we have only one such point.
	}
 $t_\infty$ on the covering where $z$ diverges as $z \approx b_{t_\infty} (t-t_\infty)^{-n_{t_\infty}}$; here $n_{t_\infty} = 1$ and $n_\infty = n$.

Expanding $z(t)$ around  $t=0,\infty,t_1,x,t_\infty$, and taking into account (\ref{tis}), we can read the necessary parameters
\be
\begin{split}	\label{biss}
b_0 &= x^{-1}(x-1)^{-n}(x+n)^{n+1}(x+n-1)^{-n},
\\
b_{\infty} &= (x-1)^{-n-1}(x+n)^n(x+n-1)^{-n+1},
\\
b_{t_1}&= -n(x-1)^{-2}(x+n)^2(x+n-1)^{-2}(x+ \tfrac{n-1}{2}),
\\
b_x &=x^{n-3}(x-1)^{-n-1}(x+n)^{n+1}(x+n-1)^{-n+1}(x+ \tfrac{n-1}{2}) ,
\\
b_{t_\infty} &= n x^{-1} (x-1)^{-n-1} (x+n)^{-1} (x+n-1)^{-n} .
\end{split}
\ee
Note that the coefficient at $t = t_0$ is not necessary, since $z \approx b_{t_0} ( t-t_0)$ has a trivial monodromy, hence there is no Liouville contribution at this point. Inserting (\ref{biss}) into the Liouville action we find
\begin{align}	\label{Liouvx}
\begin{split}
S_L &= 
	-
	\frac{2 +5n ( n -1)}{4n} \log |x|
	+
	\frac{2 + 5n (n+1) }{4n} \log | x-1 |
\\
&\quad
	+
	\frac{2 - n ( n+1)}{4n} \log |x+n|
	-
	\frac{2 - n(n-1)}{4n} \log |x+n-1|
\\
&\quad
	-
	\frac{1}{2} \log |x + \tfrac{n-1}{2} |
\\
&\quad
	- \log 2 + \frac{1}{2}  \log n 
	+ \text{Regulation terms}	
\end{split}
\end{align}
The numerical terms in the last line are normalization-dependent, and can be absorbed in the definition of $\s_n$.%
		\footnote{%
		In a sense, the LM technique really gives a path-integral \emph{definition} of twist operators, through the covering map and the insertion of regular (``vacuum'') patches at the circles cut off from the covering surface in the regularization procedure.
		}
The dynamical part of the four-point function $e^{S_L}$ is given by the three first lines, parameterized by $x$.
As expected, the result is the same as found in \cite{Lima:2020kek} via the stress-tensor method, cf.~\S\ref{SectStressTens} below.

As said, the bosonic and fermionic factors in (\ref{GsBF}) are computed from the untwisted theory living on the covering surface, and are also naturally parameterized by $x$.  The fermions appear in (\ref{gu}) as exponentials  inserted at branching points (\ref{coefb}). As shown by LM \cite{Lunin:2001pw}, an exponential operator lifts to the covering surface as
\be\label{rim}
e^{ip(\phi^1-\phi^2)}(z_*) \mapsfrom b_*^{-{p^2 / n_*}}e^{ip(\phi^1-\phi^2)}(t_*) .
\ee
At  $t = \infty$, the coefficient at the r.h.s.~is instead $({1/ b_*})^{-{p^2 / n_*}}$, obtained by mapping $t \mapsto 1/ t'$, then taking $t'=0$.
Thus the Ramond fields $R^{0\pm}_{[n]}(z , \bar z)$ lift to the covering as
\be	\label{R0lift}
\begin{split}
R_n^{0\pm} (0,\bar 0) 
	&\mapsfrom b_{0}^{-{1\over 4n}}
			e^{\pm {i\over 2}(\phi^1-\phi^2)}(0)\times c.c
\\
R_n^{0\pm} (\infty,\bar \infty) 
	&\mapsfrom b_{\infty}^{{1\over 4n}}
			e^{\pm {i\over 2}(\phi^1-\phi^2)}(\infty)\times c.c
\end{split}
\ee
and the fermionic exponentials in the interaction operators lift as
\be\label{mapoint}
b_{t_1}^{-{1\over 8}}e^{\pm {i\over 2}(\phi^1-\phi^2)}(t_1) \times c.c.
\qquad
\text{and}
\qquad
b_{x}^{-{1\over 8}}e^{\pm {i\over 2}(\phi^1-\phi^2)}(x) \times c.c.
\ee
When looking at the product of interaction terms, we note that, since they are inside correlation functions, only terms multiplied by the self-conjugate combinations $\pa X^{a\dagger} \pa X^a$ and $\pa X^a \pa X^{a\dagger}$ do not vanish. 
The product of interaction operators lifted to the covering surface, 
\be	\label{liftofOint}
\Oint(1,\bar 1) \Oint(u,\bar u) \mapsfrom \Oincov (t_1,\bar t_1) \Oincov(x, \bar x) = I+II+III+IV
\ee
can then be organized as a sum of four terms, respectively
\begin{align}
\begin{split}		 \label{ints}
&
I \sim 
	e^{-{i\over 2}(\phi^1-\phi^2)}
	\pa X^{2\dagger}(t_1)
	e^{{i\over 2}(\phi^1-\phi^2)}
	\pa X^2(x)
\\
& \times 
	\bigg( 
	e^{{i\over 2}(\bphi^1-\bphi^2)}
	\bar \pa X^2(\bar t_1)
	e^{-{i\over 2}(\bphi^1-\bphi^2)}
	\bar \pa X^{2\dagger}(\bar x)
	+
	e^{-{i\over 2}(\bphi^1-\bphi^2)}
	\bar\pa X^1(\bar t_1)
	e^{{i\over 2}(\bphi^1-\bphi^2)}
	\bar\pa X^{1\dagger}(\bar x) 
	\bigg)
\\
&II \sim 
	e^{{i\over 2}(\phi^1-\phi^2)}
	\pa X^2(t_1)
	e^{-{i\over 2}(\phi^1-\phi^2)}
	\pa X^{2\dagger}(x)
\\
&\times 
	\bigg( 
	e^{-{i\over 2}(\bphi^1-\bphi^2)}
	\bar\pa X^{2\dagger}(\bar t_1)
	e^{{i\over 2}(\bphi^1-\bphi^2)}
	\bar\pa X^2(\bar x)
	+
	e^{{i\over 2}(\bphi^1-\bphi^2)}
	\bar\pa X^{1\dagger}(\bar t_1)
	e^{-{i\over 2}(\bphi^1-\bphi^2)}
	\bar\pa X^1(\bar x) 
	\bigg)
\\
&III \sim 
	e^{{i\over 2}(\phi^1-\phi^2)}
	\pa X^{1\dagger}(t_1)
	e^{-{i\over 2}(\phi^1-\phi^2)}
	\pa X^1(x)
\\
& \times 
	\bigg( 
	e^{-{i\over 2}(\bphi^1-\bphi^2)}
	\bar\pa X^1(\bar t_1)
	e^{{i\over 2}(\bphi^1-\bphi^2)}
	\bar\pa X^{1\dagger}(\bar x)
	+
	e^{{i\over 2}(\bphi^1-\bphi^2)}
	\bar\pa X^2(\bar t_1)
	e^{-{i\over 2}(\bphi^1-\bphi^2)}
	\bar\pa X^{2\dagger}(\bar x) 
	\bigg)
\\
&IV \sim 
	e^{-{i\over 2}(\phi^1-\phi^2)}
	\pa X^1(t_1)
	e^{{i\over 2}(\phi^1-\phi^2)}
	\pa X^{1\dagger}(x)
\\
&\times 
	\bigg( 
	e^{{i\over 2}(\bphi^1-\bphi^2)}
	\bar\pa X^{1\dagger}(\bar t_1)
	e^{-{i\over 2}(\bphi^1-\bphi^2)}
	\bar\pa X^1(\bar x)
	+
	e^{-{i\over 2}(\bphi^1-\bphi^2)}
	\bar\pa X^{2\dagger}(\bar t_1)
	e^{{i\over 2}(\bphi^1-\bphi^2)}
	\bar\pa X^2(\bar x) 
	\bigg)
\end{split}
\end{align}
We note that, to obtain the correct signs in the above expressions, we must insert the proper cocycles in the bosonization of  $\psi^a$ and $\psi^{a\dagger}$,  see \cite{Burrington:2012yq,Burrington:2015mfa}.
We have ignored multiplicative factors coming from the lifting, given in (\ref{mapoint}). These factors are crucial, and we will carefully restore them later.

The $\pa X$s in (\ref{ints}) give the bosonic contribution. It is not hard to see from their structure that all four terms give the same, very simple contribution, namely products of one holomorphic and one anti-holomorphic two-point function of bosonic currents:
\be	\label{Gbosn}
\begin{split}
G_B &= 4 \times 2 \times 
	\big\langle \pa X^\dagger (t_1) \pa X (x) \big \rangle 
	\times
	\big\langle \bar \pa X^\dagger (\bar t_1) \bar \pa X(\bar x) \big \rangle
\\
	&= 8 | (t_1 - x)^{-2} |^2 
\\
	&= \tfrac12  \big| (x+n)^{2} (x + \tfrac{n-1}{2})^{-2} \big|^2 ,
\end{split}	
\ee
where here $\pa X$ is one of the bosonic currents $\pa X^a$.
In the last line, we have used (\ref{tis}).
The numerical factor of $2\times 4$ comes from the two contributions in each of the four terms. Clearly, $G_B$ really factorizes as in (\ref{GsBF}). 
Note that the bosonic currents do not carry factors of $b$ when lifted.

The fermionic contributions to the terms (\ref{ints}) 
are more complicated, but can be reduced to a basic correlation of exponentials. The holomorphic fermionic contribution from the term $I$, apart from the $b$ factors, is 
\be	\label{Expsnuethol}
\begin{split}
	\Big\langle 
	e^{-{i\over 2}(\phi^1-\phi^2)}(\infty)
	e^{-{i\over 2}(\phi^1-\phi^2)}(t_1)
	e^{{i\over 2}(\phi^1-\phi^2)}(x)
	e^{{i\over 2}(\phi^1-\phi^2)}(0)
	\Big\rangle
	=
	(t_1-x)^{-{1\over 2}} \left( { x / t_1} \right)^{{1\over 2}}
\end{split}
\ee
while the anti-holomorphic part of $I$ gives
\be	\label{ExpsnuetAnhol}
\begin{split}
\Big\langle 
e^{{i\over 2}(\bphi^1 - \bphi^2)}(\bar\infty)
&	\Big[
	 e^{{i\over 2}(\bphi^1-\bphi^2 )}(\bar t_1)
	 e^{-{i\over 2}(\bphi^1-\bphi^2)}(\bar x)
\\
&
\qquad\qquad
	 +
	 e^{-{i\over 2}(\bphi^1-\bphi^2)}(\bar t_1)
	 e^{{i\over 2}(\bphi^1-\bphi^2)}(\bar x) 
	 \Big]
e^{-{i\over 2}(\bphi^1-\bphi^2)}(\bar 0)
\Big\rangle
\\
&
= 
(\bar t_1-\bar x)^{-{1\over 2}}
\Big[  \left( {\bar x / \bar t_1} \right)^{{1\over 2}} 
	+
	\left( {\bar t_1 / \bar x} \right)^{{1\over 2}} 
\Big] .
\end{split}
\ee
The term $IV$ gives exactly the same contribution.
The terms $II$ and $III$ both give, also, equal contributions, which are slightly different from the one above: the holomorphic terms are now
\be
\begin{split}
\Big\langle 
	e^{-{i\over 2}(\phi_1-\phi_2)}(\infty)
	e^{{i\over 2}(\phi_1-\phi_2)}(t_1)
	e^{-{i\over 2}(\phi_1-\phi_2)}(x)
	e^{{i\over 2}(\phi_1-\phi_2)}(0)
\Big\rangle
	=
	(t_1-x)^{-{1\over 2}} \left( {t_1/ x} \right)^{{1\over 2}}
\end{split}
\ee
while the anti-holomorphic part turns out the same as before,
\be	\label{Expsnuethol19}
\begin{split}
\Big\langle 
e^{{i\over 2}(\bphi_1-\bphi_2)}(\bar\infty)
&	\Big[ 
	e^{-{i\over 2}(\bphi_1-\bphi_2)}(\bar t_1)
	e^{{i\over 2}(\bphi_1-\bphi_2)}(\bar x)
\\
&
\qquad\qquad
	+
	e^{{i\over 2}(\bphi_1-\bphi_2)}(\bar t_1)
	e^{-{i\over 2}(\bphi_1-\bphi_2)}(\bar x) 
	\Big]
e^{-{i\over 2}(\bphi_1-\bphi_2)}(\bar0)
\Big\rangle
\\
&
= 
	(\bar t_1-\bar x)^{-{1\over 2}}
	\Big[
		\left( {\bar t_1/ \bar x} \right)^{{1\over 2}}
		+
		\left( {\bar x / \bar t_1} \right)^{{1\over 2}} 
	\Big]
\end{split}
\ee
Combining $I+II+III+IV$, the full fermionic part of $G_0(u,\bar u)$ is therefore
\be
\begin{split}
G_F &= 
	\Big|  
	b_0^{-{1\over 4n}} 
	b_\infty^{{1\over 4n}} 
	b_1^{-{1\over 8}} 
	b_x^{-{1\over 8}} 
	(t_1-x)^{-{1\over 2}} 
	\Big[
	\left( { x / t_1} \right)^{{1\over 2}}
	+
	\left( { t_1/ x} \right)^{{1\over 2}}
	\Big] 
	\Big|^2 
\\
	&=
	\Big|
	2^3
	x^{ \frac{2-n(n+1)}{8n} }
	(x-1)^{- \frac{2-n(n-1)}{8n} }
	(x+n)^{- \frac{2+n(n+3)}{8n} }
	(x+n-1)^{ \frac{2+n(n-3)}{8n} }
\\
&\qquad\qquad\qquad\qquad\qquad\qquad
	\times
	(x+ \tfrac{n-1}{2})^{-\frac34}	
	\Big[
	x (x + n -1)
	-
	\tfrac{n-1}{2}
	\Big] 	
	\Big|^2	
\end{split}
\ee
We have restored the factors of $b_{t_1},b_x$ coming from  lifting the exponentials in $\Oint$, and the factors of $b_0,b_\infty$ coming from lifting the Ramond fields. Their powers are dictated by Eq.(\ref{rim}).
To obtain the final expression for the four-point function, we must be very careful and recall that when lifting $\Oint$ to the covering, there is an additional factor of
\be
\big| b_{t_1}^{-1/2} b_x^{-1/2} \big|^2	\label{addbfac}
\ee
coming from the Jacobian of the contour integrals defining $\Oint$ as the action of super-current modes on a chiral field.
Combining all the factors above, we finally obtain the complete function (\ref{GsBF}) as
\be\label{r0fun}
\begin{split}
G_0(x,\bar x)
	=
	\Big| 
	C 
	x^{-{5n\over 4}+2}
	(x-1)^{{5n\over 4}+2}
	(x+n)^{-{3n\over 4}}
	(x+n-1)^{{3n\over 4}}
	(x+ \tfrac{n-1}{2})^{-4}
\\	
	\times
	\Big[ 
	x( x + n-1 )
	+ 
	\tfrac{1-n}{ 2}
	\Big]
	\Big|^2 .
\end{split}
\ee
We have grouped factors of 2 and of $n$ inside an overall constant $C$ which depends on the normalization of the fields. It will be determined to be
\be
C = \frac{1}{16n^2} 	\label{Cn}
\ee
in Sect.~\ref{OPEsingle} below.
Let us note that the corresponding function for R-charged fields, which we have derived in \cite{Lima:2020kek,Lima:2020boh} using the stress-tensor method, is computed with the LM technique in Appendix \ref{SectLMRcharged}.

In Eq.(\ref{r0fun}),  the four-point function has been written completely in terms of $(x,\bar x)$, which is the pre-image of $(u,\bar u)$ on the covering surface.
 This is achieved after writing explicitly all the $b$s, as well as $t_1$, etc., according to Eqs.(\ref{tis}) and (\ref{biss}).
To find $G_0(u,\bar u)$ from $G_0(x,\bar x)$, we have to invert the function $u(x)$ in Eq.(\ref{ux}). In general, there is a collection of $2n$ inverses $x_{\frak a}(u)$, which are related to the possible configurations between the permutation cycles entering the twists. Since $G_0(u,\bar u)$ is  a sum over all orbits of the cycles, every inverse contributes, and 
\be
G_0(u,\bar u)= \sum_{\frak a = 1}^{2n} G_0(x_{\frak a}(u) , \bar x_{\frak a}(\bar u)).
\ee
See e.g.~\cite{Lima:2020kek} and references therein for a detailed discussion of this point.

\subsection{The stress-tensor method}	\label{SectStressTens}

A second way of computing the function $G_0(x,\bar x)$ is the stress-tensor method \cite{Dixon:1986qv,Arutyunov:1997gt,Pakman:2009zz}, which we have recently implemented in the derivation  of an analogous function for R-charged Ramond fields \cite{Lima:2020boh,Lima:2020kek}.
The central idea is to solve a first-order differential equation resulting from the conformal Ward identity:
\be
\pa_u \log G_0(u,\bar u) = \underset{z = u}{ \rm{Res} } \, F(z), \label{STdif1}
\ee
where
\be
F(z) = 
	\frac{
	\big\langle 
	T(z)
	R^{0-}_{[n]}(\infty,\bar \infty) \Oint(1,\bar 1) \Oint (u , \bar u) R^{0+}_{[n]}(0,\bar 0) 
	\big\rangle
	}{G_0(u,\bar u)}	.
\ee
We are again faced with the difficult monodromies, so instead of finding $F(z)$ we calculate the corresponding function $F_{\rm{cover}}(t)$, obtained after insertion of the stress-tensor $T(t)$ into the correlator of the lifted operators on the covering surface. 
The stress-tensor on the covering is given by (\ref{stres}) without the sum over copies.  The Ramond fields and the interaction operator lift to the covering as in (\ref{R0lift}) and (\ref{ints}). 
An advantage of the stress-tensor method is that the overall factors of $b$, crucial in the LM technique, are irrelevant here, as they cancel in the fraction. 
So let us denote by 
\be
S^{0\pm} = e^{\pm \frac{i}{2} ( \phi^1 - \phi^2)}
\ee
the covering-surface spin fields corresponding to the lifted Ramond fields $R^{0\pm}_{n}$. We find
\begin{align}
\begin{split}
F_{\rm{cover}} 
&=
\frac{
	\big\langle T(t) S^{0-}(\infty, \bar \infty) \Oincov(t_1,\bar t_1) \Oincov (x , \bar x) S^{0+}(0,\bar 0) \big\rangle
	}{
	\big\langle S^{0-}(\infty,\bar \infty) \Oincov(t_1,\bar t_1) \Oincov (x , \bar x) S^{0+}(0,\bar 0) \big\rangle
	} 
\\
&=
	\frac{(t_1 - x)^2}{(t-t_1)^2 (t - x)^2} 
\\
&\quad
	-\frac{1}{4} \Bigg[ 
	\frac{1}{t^2} + \left( \frac{1}{t-t_1} - \frac{1}{t-x} \right)^2
	- \frac{2}{t(t-t_1)}
	-\frac{2}{t(t-x)}
\\
&\qquad\qquad
	+ \frac{4}{t(t-t_1)} 
	\frac{
	\big\langle S^{0-}(\infty,\bar \infty) V_+(t_1,\bar t_1) V_- (x,\bar x) S^{0+}(0,\bar 0) \big\rangle	
		}
		{
	\big\langle S^{0-}(\infty,\bar \infty) \Oincov(t_1,\bar t_1) \Oincov (x , \bar x) S^{0+}(0,\bar 0) \big\rangle
		}
\\		
&\qquad\qquad
	+ \frac{4}{t(t-x)} 
	\frac{
	\big\langle S^{0-}(\infty,\bar \infty) V_-(t_1,\bar t_1) V_+(x,\bar x) S^{0+}(0,\bar 0) \big\rangle	
		}
		{
	\big\langle S^{0-}(\infty,\bar \infty) \Oincov(t_1,\bar t_1) \Oincov (x , \bar x) S^{0+}(0,\bar 0) \big\rangle
		}
		\Bigg]
\end{split}\label{TeSSOOOO}
\end{align}
(Note that, on covering-surface correlators, we remove the twist label of $\Oint$.)
The expression in the first line is the bosonic part of  $F_{\rm{cover}}$, coming from contractions of $\pa X^a$ and $\pa X^{a\dagger}$; it does not depend on the Ramond fields, and is the same as the one found in \cite{Lima:2020boh,Lima:2020kek}. The remaining terms come from fermionic contractions between the normal-ordered $\colon\pa\phi^a\pa\phi^a\colon$ on $T(t)$ and the exponentials in the other operators. 
The operators $V_\pm$ are part of the interaction operator on the covering surface, 
\be
\Oincov(t,\bar t) = a \Big[ V_+ (t,\bar t) + V_-(t,\bar t) \Big] ;
\ee
$a$ is the appropriate combination of $b$ coefficients used in \S\ref{SectLMtech}, which are unimportant here, and 
\bsub\begin{align}
\begin{split}
V_+ (t,\bar t) =
	e^{+ \frac{i}{2} (\phi^1- \phi_2 )}
&	\Big[ 
		\pa X^{2\dagger} 
		\Big(
		\bar \pa X^1
		e^{\frac{i}{2} ( \bphi^1 - \bphi^2 )}
		-
		\bar \pa X^2
		e^{-\frac{i}{2} ( \bphi^1 - \bphi^2 )}
		\Big)
\\
&
		-
		\pa X^{1} 
		\Big(
		\bar \pa X^{1\dagger}
		e^{-\frac{i}{2} ( \bphi^1 - \bphi^2 )}
		+
		\bar \pa X^{2\dagger}
		e^{\frac{i}{2} ( \bphi^1 - \bphi^2  )}
		\Big)
	\Big]
\end{split}
\\
\begin{split}
V_- (t, \bar t)=
	e^{- \frac{i}{2} (\phi^1- \phi_2 )}
&	\Big[ 
		\pa X^{2} 
		\Big(
		\bar \pa X^{1\dagger}
		e^{- \frac{i}{2} ( \bphi^1 - \bphi^2 )}
		+
		\bar \pa X^{2\dagger} 
		e^{\frac{i}{2} ( \bphi^1 - \bphi^2 )}
		\Big)
\\
&
		+
		\pa X^{1\dagger} 
		\Big(
		\bar \pa X^{1}
		e^{\frac{i}{2} ( \bphi^1 - \bphi^2 )}
		-
		\bar \pa X^{2}
		e^{-\frac{i}{2} ( \bphi^1 - \bphi^2  )}
		\Big)
	\Big] .
\end{split}
\end{align}\esub
The terms containing products of $V_\pm$ in (\ref{TeSSOOOO}) are absent in the analogous computation with R-charged Ramond fields $R^{\pm}_{[n]}$ detailed in \cite{Lima:2020kek}, because of a cancellation of factors peculiar to that case.
This simplification can also be seen in the LM technique computation, as we show in Appendix \ref{SectLMRcharged}.
Direct computation of the correlators in the last lines of (\ref{TeSSOOOO}) leads again to the four-point functions of exponentials found in \S\ref{SectLMtech}, and we finally obtain
\be
\begin{split}
F_{\rm{cover}} (t) 
&=
 	\frac{(t_1 - x)^2}{(t-t_1)^2 (t - x)^2} 
\\
&\quad
+
\frac{1}{4} \Bigg[
	\left(	\frac{1}{t-t_1}	-	\frac{1}{t-x}	\right)^2
	+
	\frac{1}{ t^2} + \frac{2(t_1-x)^2}{t(t-t_1)(t-x)(t_1+x)} \Bigg] .
\end{split}
\label{FcovernutRam}
\ee

The next step in the stress-tensor method is to map this function back to the base, taking into account the Schwarzian derivative $\{t,z\}$ in the anomalous transformation of $T$, which actually accounts for the twists' contribution, playing the role of the Liouville factor in the LM technique, but without requiring any regularized normalization of the twist field  $\s_n$.%
		\footnote{%
		The solution of (\ref{Tmethx}) with $F_{\rm{cover}} = 0$ is precisely the function $G_\s = e^{S_L}$ with $S_L$ given in Eq.(\ref{Liouvx}), see \cite{Lima:2020kek}.
		}
Again, what we get is an expression fully parameterized by $x$, as it is already clear from (\ref{FcovernutRam}). So, instead of solving (\ref{STdif1}), we make a change of variables, to solve
\bsub\label{Tmethx}\begin{flalign}
&& & \pa_x  \log G_0(x) = u'(x) H_0(x) &&		\label{STdif2}
\\
\text{where}
&&
& H_0(x) = \underset{z = u}{ \rm{Res} } \, F(z) 
	= 2 \, \underset{z = u}{ \rm{Res} } \, \Big[ \tfrac{1}{2} \{ t, z \} + \left( dt/dz \right)^2 F_{\rm{cover}} (t(z)) \Big] . &&
\end{flalign}\esub
The factor of 2 comes from the sum over the two copies involved in the twist $\s_2$ around $z = u$. Here $t(z)$ is any of the two local inverses of $z(t)$ near $z = u$. Details of the inverse map can be found in \cite{Lima:2020kek}.
The function $H_0(x)$ is rational, and integration of (\ref{STdif2}) gives the expression inside the absolute value bars in Eq.(\ref{r0fun}), with $C$ as the integration constant. Of course, $G_0(x,\bar x) = G_0(x) G_0(\bar x)$, where $G_0(\bar x) = \overline{G_0(x)}$ is found with the same procedure, but carried on with the anti-holomorphic stress-tensor $\tilde T(\bar t)$.

\section{OPE channels}	\label{OPEsingle}

The behavior of $G_0(u,\bar u)$ near the singular points $u=0,1,\infty$ give the OPE channels of the fields involved in the correlation function. 
As explained before, for each of these points there is a collection of distinct limits $x_{\frak a}(u)$, related to the different possibilities of combining  permutations in the conjugacy class defined by the twists. 
Each of these different limits of $x$ will therefore give a different OPE channel corresponding to operators in distinct twisted sectors.

Let us start with the limit $u\to 1$, the OPE of two interaction operators. Examining the explicit expression (\ref{ux}) we see that there two channels, $x \to \infty$ and $x \to \frac{1-n}{2}$. In the former, 
\be
G_0(u,\bar u)|_{x \to \infty} =  \frac{|16n^2 C|^2}{|1-u|^4} + \text{non-sing.}
\ee
The powers of $u$ imply that this expression corresponds to the two-point function of the interaction operator. We first notice that since the latter, as well as the two-point function of Ramond fields, are normalized to one, then $C$ is indeed fixed to the value (\ref{Cn}). Second, there is no subleading term of order $|1-u|^{-2}$, which would correspond to a field of dimension one. So there is no such field in the OPE of the interaction fields, confirming it is a truly marginal deformation. 
In the other channel $x \to \frac{1-n}{2}$,
\begin{align}	\label{Channs3}
G_0(u,\bar u)|_{x \to \frac{1-n}{2}} =
	 \frac{
	 \big|
	 2^2
	 3^{\frac43} 
	 (n -1)^{\frac{2-3n}{6} }
	 (n+1)^{ \frac{2+3n}{6} }
	 n^{ -\frac23 } 
	 \big|^2
	}
	{|1-u |^{8/3}} 
	 + \rm{O} (|1-u|^{-\frac43} )  .
\end{align}
The leading singularity corresponds to the twist field $\sigma_3$ with dimension $\Delta^\s_n = \frac43$, and its coefficient gives the product of the structure constants of the interaction fields and the Ramond fields $R^{0\pm}$ with the twist field $\sigma_3$. We notice again the absence of the subleading term $\sim |1-u|^{-2}$, that would correspond to a field of dimension one in the OPE of the interaction fields.

Let us turn now to the limit $u\to 0$ and the OPE of the interaction and the Ramond fields, $\Oint(u,\bar u)R^{0+}_{[n]}(0)$.
From (\ref{ux}), it follows that there are again two channels, $x \to 0$ and $x \to -n$,
\begin{align}
G_0(u,\bar u)|_{x \to 0} &= 
	\frac{
	\big|
	 2^{-1}
	 (n -1)^{ - \frac{n-2}{2}}
	 n^{ \frac{n^2-4n}{2(n-1)} }
	 |^2
	 }{
	|u|^{ \frac{5n - 8}{4(n-1)}} 
	}
	\Big| 
	1 + \rm{const.} \ u^{\frac{1}{n-1}}  +\cdots
	\Big|^2
\\		\label{ChannYp}
G_0(u,\bar u)|_{x \to -n} &=
	\frac{
	\big|
	 2^{-1}
	 (n + 1)^{ \frac{n-2}{2}}
	 n^{ - \frac{n^2+4n}{2(n+1)} }
	 \big|^2
	 }{
	|u|^{ \frac{3n}{4(n+1)}} 
	}
	\Big| 
	1 + \rm{const.} \ u^{\frac{1}{n+1}} +\cdots
	\Big|^2
\end{align}
The leading singularities in the above channels reveal operators $\cal Y^{0+}_{n-1}$ and $\cal Y^{0+}_{n+1}$, respectively, in the twisted sectors with $\s_{n\pm1}$, i.e. we get the fusion rule
\be	\label{OPER0pm}
[ \Oint ] \times [ R^{0+}_{[n]} ]  = [ \cal Y^{0+}_{[n-1]}] +  [ \cal Y^{0+}_{[n+1]} ] . 
\ee
We can read the dimension of $\cal Y^{0+}_{[m]}$ to be
$\Delta^{\cal Y}_m = h^{\cal Y}_m + \tilde h^{\cal Y}_m$ with the weights
\be	\label{hclY}
h^{\cal Y}_m 
= 
 \frac{m}{4} + \frac{3}{4m}
= \tilde h^{\cal Y}_m,
\ee
where the dimension of the twist field is given in (\ref{twistdim}).
By charge conservation,  $\cal Y^{0+}_{[m]}$ is R-neutral and part of a doublet of the internal SU(2) symmetry; the second field in the doublet, $\cal Y^{0-}_{[m]}$, can be found by taking the corresponding OPE limit for $\Oint  R^{0-}_{[n]}$.%
		\footnote{%
		This requires bringing $R^{0-}_{[n]}$ from infinity by a conformal transformation of the four-point function (\ref{gu}), i.e.~fixing the points in (\ref{confor}) differently, but note that this gives the conjugate of  same function we have analyzed, hence the same dimensions, etc.
		}

The leading coefficients in the expansions (\ref{Channs3})-(\ref{ChannYp}) are the structure constants of the operators involved in the respective OPE channels, which can be expressed in terms of three-point functions:
\begin{align}	\label{Rs3R}
&
\Big\langle 
R^{0-}_{n}(\infty, \bar \infty) 
\s_{3}(1,\bar 1)
R^{0+}_{n}(0, \bar 0) 
\Big\rangle
=
	 2^{-\frac52}
	 3^{-\frac43} 
	 (n -1)^{\frac{2-3n}{3} }
	 (n+1)^{ \frac{2+3n}{3} }
	 n^{ -\frac43 } 
\\
&
\Big|
\Big\langle
R^{0\mp}_{n}(\infty, \bar \infty) 
\Oincov_{2}(1,\bar 1)
{\cal Y}^{0\pm}_{n-1}(0, \bar 0) 
\Big\rangle
\Big|^2
=
	 2^{-2}
	 (n -1)^{ 2-n}
	 n^{ \frac{n^2-4n}{n-1} }
\\
&
\Big|
\Big\langle
R^{0\mp}_{n}(\infty, \bar \infty) 
\Oincov_{2}(1,\bar 1)
{\cal Y}^{0\pm}_{n+1}(0, \bar 0) 
\Big\rangle
\Big|^2
=
	 2^{-2}
	 (n + 1)^{ n-2}
	 n^{ - \frac{n^2+4n}{n+1} }
\end{align}
To obtain (\ref{Rs3R}) we have used the structure constant 
$\langle \Oincov_{2}  \s_{3} \Oincov_{2} \big\rangle = 2^{\frac{13}{3}} 3^{4}$, derived in \cite{Lima:2020kek}.
These expressions give the values of structure constants of operators whose twists are one representative of their conjugacy classes, i.e.~there is no sum over orbits; see the discussion in \cite{Lima:2020kek}.

\bigskip

\noindent
{\bfseries Chiral fields}

\noindent
We can compare the operators ${\cal Y}^{0\pm}_n$ found above to the ones that appear in the OPE between $\Oint$ and the chiral fields $O^{(1,1)}_n$. 
As discussed in \cite{Burrington:2012yq,deBeer:2019ioe}, computing four-point functions like $G_0(u,\bar u)$ and then taking the coincidence limits is a very efficient way of finding only the non-vanishing structure constants, as well as the set of fields appearing in the OPE of an operator with the deformation.
For the middle-cohomology chiral fields, the functions we need are
\be
{\scr G}_{\pm}(u, \bar u)
	\equiv
\Big\langle [ O^{(1,1)\pm}_{[n]} ]^\dagger(\infty, \bar \infty) \Oint(1,\bar 1) \Oint(u,\bar u) O^{(1,1)\pm}_{[n]}(0,\bar 0) \Big\rangle . \label{4ptCh}
\ee
The same methods used in Sect.\ref{SectTwoPointFuncts} can be directly applied here. The covering map and the Liouville factor (\ref{Liouvx}) are the same, and so is the bosonic factor (\ref{Gbosn}), since there are no new bosons in (\ref{4ptCh}) when compared to (\ref{gu}). The only difference is in the fermionic contractions on the covering, e.g.~(\ref{Expsnuethol})-(\ref{Expsnuethol19}), which now involve slightly different coefficients in the exponentials corresponding to the bosonized expression for $O^{(1,1)\pm}_n$, cf.~Eq.(\ref{O11pm}).
(Alternatively, formula (\ref{FcovernutRam}) is changed.)
Note that $[O^{(1,1)\pm}_n]^\dagger \neq O^{(1,1)\mp}_n$, but nevertheless it has charges $\frak j^3 = \mp \frac12$ (and is R-neutral).
Hence the two functions ${\scr G}_\pm$ are not the complex conjugate of one another, and both are allowed by charge conservation.
It turns out, however, that the two functions are equal, ${\scr G}_+ = {\scr G}_-  = {\scr G}$, where, with the parameterization in terms of $x$ as before,
\be	\label{Go11x}
\begin{split}
{\scr G} (x,\bar x)
	=
	\Big| 
C \, \frac{
		x^{ 2-n }
		(x-1)^{ 2+n }
		(x+n)^{ -n }
		(x+n-1)^{ n }
		}{
		(x + \frac{n-1}{2})^4
		}
		\Big[ x ( x+n-1) - \tfrac{n-1}{2}  \Big]
			\Big|^2 .
\end{split}
\ee
From here, one can proceed to compute the OPEs. The channels from $u \to 0$, $u \to 1$ are the same functions of $x$ (recall the covering map is the same). The OPE of the two interactions as $u \to 1$ are the same as before, as expected, and fix $C = 1/16n^2$ again. Now for $u \to 0$ we find
\begin{align}
{\scr G} (u,\bar u) |_{x \to 0}
	&= 
	\frac{
		| 2^{-1} (n-1)^{-1} n^{-\frac{2n}{n-1}} |^2
		}{
		|u|^{\frac{2(n - 2)}{n-1}}
		} 
		\Big| 1 + \rm{const.} \ u^{\frac{1}{n-1}} 
				+ \cdots 
		\Big|^2
\\
{\scr G} (u,\bar u) |_{x \to -n}
	&= 
	\frac{
	| 2^{-1} (n+1)^{-1} n^{-\frac{2n}{n+1}} |^2
	}{
	| u |^{\frac{2n}{n+1}}
	} 
	\Big| 1 + \rm{const.} \ u^{\frac{1}{n+1}} 
				+ \cdots 
	\Big|^2	
\end{align}
The two channels give operators with twists $n-1$ and $n+1$, respectively, in an OPE
\be	\label{OPEO11}
[\Oint] \times [O^{(1,1) \pm}_{[n]}] = {\cal A}^{0\pm}_{[n-1]} + {\cal B}^{0\pm}_{[n+1]}
\ee
where the conformal weights of the $m$-twisted fields ${\cal A}^{0\pm}_{[m]}$ and ${\cal B}^{0\pm}_{[m]}$ can be read from the power of $u$,
\be	\label{hclAB}
h^{\cal A}_m
	= 
	\frac{m}{2} + \frac{1}{m} + \frac12
	=
\tilde h^{\cal A}_m 
	,
\qquad
h^{\cal B}_m 
	=
	\frac{m}{2} + \frac{1}{m} - \frac12
	=
	\tilde h^{\cal B}_m . 
\ee
The operators have the same charges as $O^{(1,1)\pm}_m$, namely 
$j^3 = \frac12 m$ and $\frak j^3 = \pm \frac12$.
The coefficients of the leading-order terms give the three-point functions
\begin{align}	
&
\Big|
\Big\langle
[O^{(1,1)\pm}_{n} ]^\dagger (\infty, \bar \infty) 
\Oincov_{2}(1,\bar 1)
{\cal A}^{0\pm}_{n-1}(0, \bar 0) 
\Big\rangle
\Big|^2
	=
	2^{-2} (n-1)^{-2} n^{-\frac{4n}{n-1}}
\\
&
\Big|
\Big\langle
[O^{(1,1)\pm}_{n} ]^\dagger (\infty, \bar \infty) 
\Oincov_{2}(1,\bar 1)
{\cal B}^{0\pm}_{n+1}(0, \bar 0) 
\Big\rangle
\Big|^2
	=
	 2^{-2} (n+1)^{-2} n^{-\frac{4n}{n+1}}
 \end{align}
which are equivalent to structure constants. 
 
\bigskip 
 
\noindent
Let us make some comments.
In the r.h.s.~of the OPE (\ref{OPER0pm}), involving $R^{0\pm}_n$, we have found only one operator ${\cal Y}^{0\pm}_m$, but with the two possible twist configurations resulting of the combination of the cycles $(2)(n)$. 
By contrast, in the r.h.s.~of the OPE (\ref{OPEO11}), involving $O^{(1,1)\pm}_n$, we found two different operators, each with one of the allowed twists. 
Looking at the dimensions (\ref{hclY}) and (\ref{hclAB}) of the new-found operators, we can note that they have the structure
\begin{align}
h^{\cal Y}_m &= h^\R_m + \frac{3}{4m} \ ,
\qquad
h^{\cal A}_m
	= 
	h^{(1,1)}_m + \frac{1}{m} + \frac12
	\	,
\qquad
h^{\cal B}_m 
	=
	h^{(1,1)}_m + \frac{1}{m} - \frac12
\end{align}
The presence of the dimension of the Ramond field, $h^\R_m$, in the OPE involving $R^{0\pm}_m$, and the dimension of $O^{(1,1)\pm}_m$, $h^{(1,1)}_m = \frac12 m$, in its OPE, suggests that, in each case, the effect of the deformation operator is to excite a fractional mode over the field it acts upon. These modes must belong to neutral fields, since the charges are preserved.
Based on this, one can try to infer what are the possible explicit constructions of the operators, as was done in \cite{deBeer:2019ioe} for similar OPEs involving the lowest-weight chiral $O^{(0,0)}_n$.

The existence of two operators, ${\cal A}^{0\pm}_{[m]}$ and ${\cal B}^{0\pm}_{[m]}$, in the OPE with the chiral fields, in contrast with only one operator, ${\cal Y}^{0\pm}_{[m]}$, in the OPE with the Ramond fields, is, in itself, interesting. 
In the $Z_n$ orbifold theory with central charge $c = 6n$ defined on the $n$-twisted sector, there is a spectral flow relation between $R^{0\pm}_{[n]}$ and $O^{(1,1)\pm}_{[n]}$. Obviously, the relation is not present between the results of the OPEs of these operators with $\Oint$. 
Here, we note that the fields in the r.h.s.~of the OPEs belong to twist sectors different from the sectors of the fields in the l.h.s., but one can apply, for example, a spectral flow of the algebra with $c = 6(n-1)$ and see where ${\cal Y}^{0\pm}_{n-1}$ is mapped to: it is neither to ${\cal A}^{0\pm}_{n-1}$ nor to ${\cal B}^{0\pm}_{n-1}$.
As discussed in \cite{Lima:2020kek}, one should not expect that spectral flow relations be preserved in this way, when we make products of operators, such as $R^{0\pm}_{[n]}$ and $\Oint$, belonging to different twist sectors.
Analogous phenomena are also observed in the OPEs of the R-charged Ramond fields, $R^\pm_{[n]}$, and of the universal cohomology chirals, $O^{(1\pm1, 1\pm1)}_{[n]}$, with the interaction $\Oint$.%
	\footnote{%
	The OPE 
	$[\Oint] \times [R^\pm_{[n]}$ 
	was found in \cite{Lima:2020kek}. The OPE 
	$[\Oint] \times [O^{(0,0)}_{[n]}]$
	was found in \cite{Pakman:2009mi}; see also \cite{deBeer:2019ioe}.
	}
It would be very interesting to investigate further the nature of these operators in the r.h.s.~of the OPEs, and try to understand what is, exactly, the relation between ${\cal A}^{0\pm}_{[m]}$, ${\cal B}^{0\pm}_{[m]}$ and $R^{0\pm}_{[n]}$, etc.
This could even provide some insight into the reason why the dimensions of the chiral fields are protected to second order in $\la$-perturbation, while the Ramond fields are not, as we now show.

\section{Anomalous dimensions} \label{RenormalizationNeutral}

The second order $\la^2$-correction to the conformal dimension of the R-neutral twisted Ramond fields $R_n^{0\pm}$ is given by the integral in Eq.(\ref{correct}). 
In possession of an analytic formula for $G_0(x,\bar x)$, we can change variables from $u$ to $x$, 
\be	\label{Jndef}
J(n)  = \int \! d^2x \, | u'(x)|^2 G_0(x,\bar x) ,
\ee
and obtain a more recognizable form after a further change of variables,
\be
y(x) = -  (\tfrac{n+1}{2})^{-2}(x-1)(x+n) ,	\label{ChangofCorrd}
\ee
leading to
\begin{align}
	\begin{split}
J(n) 	&= 
	\frac{1+2n-3n^2}{64 \, n^2 (n+1)^2}
	 \int \! d^2y\, | y  |^{2a} | 1 - y |^{2b} |y - w_n|^{2c}
\\
&\quad
		+
		\frac{(n+1)^2}{256 \, n^2} 
		\int \! d^2y\, | y  |^{2a} | 1 - y |^{2b} |y - w_n|^{2(c+1)}
\\
	&\quad
		+
		\frac{n-1 }{128 n^2} \int\! d^2y\, | y  |^{2a} | 1 - y |^{2b} |y - w_n|^{2c}  \big( y +\bar y  \big)
	\end{split}	\label{lastline}
\end{align}
where we have introduced the parameters 
\be	\label{DFpara}
a= \tfrac{1}{4}n , \quad b = - \tfrac{3}{2} , \quad c = - \tfrac{1}{4} n ,
\quad w_n = \frac{4n}{(n+1)^2} .
\ee
The integral in the last line of (\ref{lastline}) vanishes. Its imaginary part is zero because $\rm{Im} (y) = - \rm{Im} (\bar y)$; meanwhile, the integrand of the real part, containing $ 2 \rm{Re}(y)$, is odd so the integral over the Real line vanishes.
We are thus left with 
\begin{align}	\label{Jdf}
	\begin{split}
J(n)
	&= 
	\frac{1+2n-3n^2}{64 \, n^2 (n+1)^2}
	 \int \! d^2y\, | y  |^{2a} | 1 - y |^{2b} |y - w_n|^{2c}
\\
&\quad
		+
		\frac{(n+1)^2}{256 \, n^2} 
		\int \! d^2y\, | y  |^{2a} | 1 - y |^{2b} |y - w_n|^{2(c+1)}
	\end{split}	
\end{align}

The integrals $\int \! d^2y\, | y  |^{2a} | 1 - y |^{2b} |y - w|^{2c}$ appearing in (\ref{Jdf}) are Dotsenko-Fateev  integrals, used as a representation of correlation functions of specific minimal models in \cite{Dotsenko:1984nm,Dotsenko:1984ad,dotsenko1988lectures}.
For the values of the parameters (\ref{DFpara}), the integrals are divergent.
Nevertheless, they can be regularized by deforming contours in the complex plane \cite{Lima:2020kek}, after which they are expressed as regular functions $I(a,b,c;w)$ of their parameters. The latter are a  combination of Gamma and regularized Hypergeometric functions ${\bf F}(\a,\b;\ga;z) \equiv F(\a,\b;\ga;z) / \Gamma(\ga)$.
Following the procedure detailed in \cite{Lima:2020kek}, we find that
\be	\label{DFanswer}
 \int \! d^2y \,  |y|^{2a} |1- y|^{2b} |y - w_n|^{2(c+q)}
=
-  \sin(\pi a) \tilde I_1^{(q)} \, I_2^{(q)} - \sin(\pi b) I_1^{(q)}  \, \tilde I_2^{(q)}
\ee
where, for $q = 0,1$, the `canonical functions' in the r.h.s.~are given by
\begin{align}
I_1^{(q)}
	&= 
	 \frac{
	2^q (1- \frac{n}{4})^q n^{2+q} \pi
	}{
	(n+1)^{2(1+q)}
	}
	F \Big( \tfrac32 , \tfrac{n+4}{4} ; 2+q ; \tfrac{4n}{(n+1)^2} \Big)
\label{I1q}
\\
I_2^{(q)}
	&= 
	\frac{
	2^{q} n^{2+q}
	(1 - \frac{n}{4})^q
	}{
	(n+1)^{2(1+q)}
	}
	\frac{\pi}{\sin(\frac{\pi n}{4})}
	F \Big( \tfrac32 , \tfrac{n+4}{4} ; 2+q ; \tfrac{4n}{(n+1)^2} \Big)
\label{I2qb}
\\
\tilde I_1^{(q)}
	&= 
	(- 2)^q \sqrt\pi  
	{\bf F} \Big( \tfrac{n}{4} - q , (-1)^q \tfrac12 ; \tfrac{n + 6 - 4q}{4} ; 1-\tfrac{4n}{(n+1)^2} \Big)
\\
\tilde I_2^{(q)} 
	&= 
	- 2 \sqrt\pi  
	\left( 1 - \tfrac{4n}{(n+1)^2} \right)^{- \frac{n}{4} - \frac12 + q}
	\Gamma ( - \tfrac{n}{4} +1+q )
	{\bf F} \Big( - \tfrac{n}{4}  , - \tfrac12 ;  \tfrac{ 2 + 4q - n}{4} ; 1-\tfrac{4n}{(n+1)^2} \Big)
\end{align}
Therefore we have
\begin{align}	\label{Jdfhyper}
	\begin{split}
J(n)
	&=  - \frac{1+2n-3n^2}{64 \, n^2 (n+1)^2}
		\Big[
		\sin(\tfrac{\pi n}{4}) \;
		\tilde I_1^{(0)}(n) \; I_2^{(0)} (n) 
		+
		I_1^{(0)}(n) \; \tilde I_2^{(0)}(n)
		\Big]
\\
&\quad
		-
		\frac{(n+1)^2}{256 \, n^2} 
		\Big[
		\sin(\tfrac{\pi n}{4}) \;
		\tilde I_1^{(1)}(n) \; I_2^{(1)} (n)
		+
		I_1^{(1)}(n) \; \tilde I_2^{(1)}(n)
		\Big]
	\end{split}	
\end{align}

The expressions above are well-defined for $n \neq 4k+4$, $k \in \mathbb N$. In this latter case, while $\sin (\pi a) = \sin (\pi c) = 0$, the expressions for $I_2^{(q)}$ and $\tilde I^{(q)}_2$ are not defined because of a pole in the Gamma functions.
	\footnote{%
	The peculiarity of this case is due to a change in the analytic properties of the DF integrals, which lose two branch cuts.}
A completely analogous situation occurs with the R-charged Ramond fields \cite{Lima:2020kek}. Now,  one can use the fact that
$
I_2^{(q)} = I_1^{(q)} / \sin( \frac{\pi n}{4} )		
$
and make a regularization of the Gamma functions in $\tilde I_2^{(q)}$, by taking $k\mapsto k - \e$ with $\e \to 0$, and extracting the finite part of a regularized DF integral 
$J(4k+4) = J^{reg}(4k+4) + \e^{-1} J^{sing}(4k+4)$,
whose finite part is 
\begin{align}	\label{JdfhyperReg}
	\begin{split}
J^{reg}(4k+4)
	&=  
	\frac{39+88 k+48 k^2}{1024 (5+9 k+4 k^2)^2}
		\ I_1^{(0)}(4k+4)
		\Big[
		\tilde I_1^{(0)}(4k+4) 
		+
		  \tilde I_2^{(0) \, reg}(k)
		\Big]
\\
&\quad
		-\frac{(5+4 k)^2}{4096 (1+k)^2}
		\ I_1^{(1)}(4k+4)
		\Big[
		\tilde I_1^{(1)}(4k+4) 
		+
		  \tilde I_2^{(1) \, reg}(k)
		\Big]
\end{split}	
\end{align}
where (see Ref.\cite{Lima:2020kek} for details) 
\be
\begin{split}
 \tilde I_2^{(q) \, reg}(k)
	&=
	\frac{
	-2 \sqrt\pi}{
	(1-\frac{16 k}{(1+4 k)^2})^{k + \frac32 - q}
	}
\\
&\quad
	\times
	\frac{(-1)^{k - q} \psi (k+1-q)}{(k-q)!}
	{\bf F} \Big(- \tfrac12 , - k-1 ; - k  + q - \tfrac12 ; 1 - \tfrac{16 k}{(1+4 k)^2} \Big)
\end{split}
\ee	
Here $\psi (k)$ is the digamma function. 
The divergent part, $J^{sing}$, is expressed as in (\ref{JdfhyperReg}), but with  $\tilde I_2^{(q) \, reg}$ replaced by 
$\tilde I_2^{(q) \, sing} = \tilde I_2^{(q) \, reg} / \psi(k+1-q)$.

\begin{figure}[t] 
\centering
	\includegraphics[scale=1]{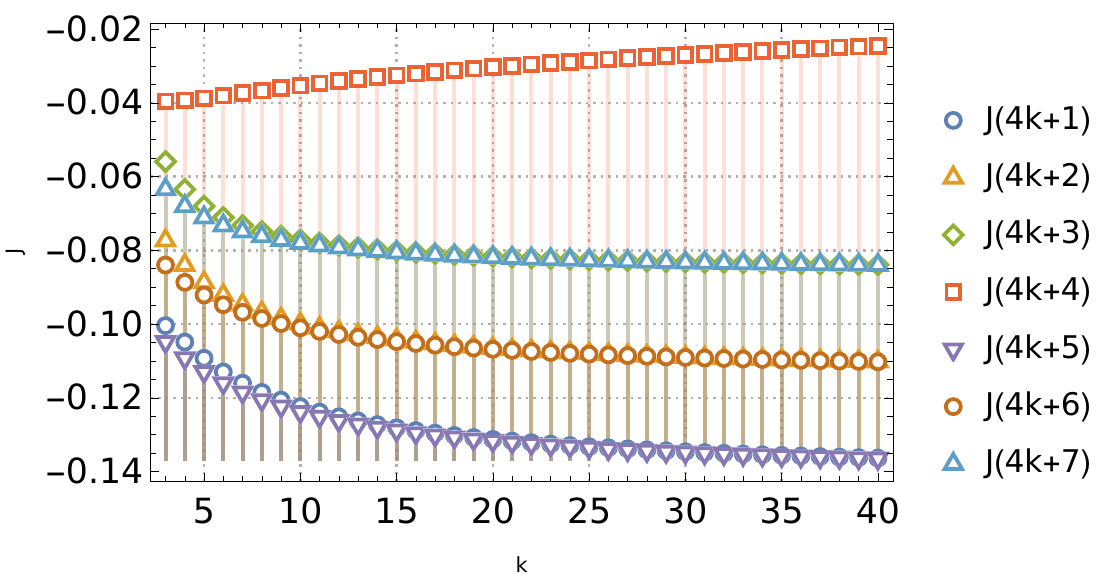}	\label{Correct_neutral_plot_UNI}
\caption{Numerical result of the integrals $J(n)$; for $n = 4k+4$, the plot corresponds to $J^{reg}(n)$.}
\label{PlotIntegrals}
\end{figure}

In Fig.\ref{Correct_neutral_plot_UNI} we plot $J(n)$ given in Eq.(\ref{Jdfhyper}) for $n\neq 4k+4$, and $J^{reg}(n)$ given by Eq.(\ref{JdfhyperReg}) when $n = 4k+4$.
Because the Gamma functions and the Hypergeometrics are (piecewise) continuous, we can see four families of ``almost continuous'' functions, distinguished by the discrete values assumed by $\sin(\frac{\pi n}{4})$.
All four families stabilize around small, negative asymptotic values for large $n$.

%
%
%
%
%

We have thus found that, after the regularization, the R-neutral twisted Ramond fields  with $n<N$ acquire an anomalous dimension at second order in perturbation theory:
\be	\label{DeltaRnla}
\Delta^R_\la(n) = \tfrac12 n + \tfrac\pi2 \la^2 |J(n)| + \cdots
\ee
where for $n = 4k+4$ we have $J^{reg}$ in the r.h.s.
Note that the integrals we have computed also give the structure constant 
$\langle R^{0-}_{[n]}(\infty) \Oint(1)R^{0+}_{[n]} \rangle = \la J(n) + \cdots$,
which vanishes in the free theory.
The fields with maximal twist $n = N$ are protected at this order of the large-$N$ expansion, since their covering surface has genus one.

\bigskip
We can perform the analogous integral using the function (\ref{Go11x}), to compute the anomalous dimension of the chiral field $O^{(1,1)\pm}_{[n]}$. 
Since the chiral field saturates a BPS bound, with $h = j^3$, we expect that it is protected under the deformation, hence the anomalous dimension vanishes. This is also expected from more general considerations \cite{Baggio:2012rr,deBoer:2008ss}. 

We can use the same change of variables (\ref{ChangofCorrd}) to reduce the integral, analogous to (\ref{Jndef}), to a Dotsenko-Fateev form (\ref{DFanswer}). But now, the exponents corresponding to (\ref{DFpara}) are $a = c = 0$, while $b = - \frac32$. Thus, instead of (\ref{Jdf}) we arrive at
\begin{align}	\label{scrJDF}
\begin{split}
{\scr J}(n) &= \int \! d^2x \, |u'(x) |^2 {\scr G}(x, \bar x)
\\
	&= \frac{1+2n-3n^2}{64 n^2 (n+1)^2} \int \! d^2y\, | 1 - y |^{-3}
		+
		\frac{(n+1)^2}{256 n^2} \int \! d^2y\,  | 1 - y |^{-3} |y - w_n|^{2}
	\end{split}	
\end{align}
The remaining integrals are again clearly divergent, and must be regularized. One can proceed just as before, but now the result is much simpler \cite{dotsenko1988lectures}
\begin{align}
\begin{split}	\label{TheDFintn1}
 \int \! d^2y \, & |y|^{2a} | y-1 |^{2b}
		=  \sin( \pi b) 
			\frac{\Gamma(1+a) 
				\Gamma^2 (1+b) 
				\Gamma (-1-a-b)
				}{
				\Gamma(-a) 
				\Gamma(2+a+b)
				} 
\end{split}
\end{align}
where we represent the integral as an analytic function of $a,b$. For the r.h.s.~of Eq.(\ref{scrJDF}), and making $u = (y-w_n)/(1+w_n)$ in the last integral, we get
\begin{align}	
\begin{split}
 \int \! d^2y \,  | y-1 |^{-3} &=  \lim_{a \to 0} \frac{4 \pi }{ \Gamma(-a) } 	= 0	,
\\
\int \! d^2y\,  | 1 - y |^{-3} |y - w_n|^{2}
	&=  (1+w_n) \lim_{a \to 1} \frac{-16\pi}{ \Gamma(-a)}
= 0	,
\end{split}
\end{align}
hence
\be
{\scr J}(n) = 0 ,
\ee
confirming that the anomalous dimension of $O^{(1,1)\pm}_n$ is protected to second order in $\la$, as predicted.

Again, this is the same qualitative result found for the R-charged Ramond fields 
$R^\pm_n$ 
in contrast with the corresponding chirals 
$O^{(1\pm1, 1\pm1)}_n$; the latter are protected while the former renormalize \cite{Lima:2020boh,Lima:2020kek}.
Since these fields are related by spectral flow of the algebra of the $n$-twisted sectors, the result might seem odd at first sight. This is a different manifestation of the phenomenon discussed at the end of Sect.\ref{OPEsingle} in relation to the OPEs --- as explained in Ref.\cite{Lima:2020kek}, the sectorial algebras with $c = 6n$ do not survive the deformation of the theory because the operator $\Oint$ mixes different twisted sectors. 
Note that the Ramond fields $R^{0\pm}_{[N]}$ with conformal weight $h^\R_N = \frac14 N = \frac{1}{24} c_{orb}$, which are related to the chirals with $h = \frac12 N = j^3$ by a spectral flow of the total orbifold algebra with $c_{orb} = 6N$, are still protected, at least in the large-$N$ limit --- but the fields with twist $n<N$ are not.

\section{Conclusion}		\label{SectConcl}

In the present paper we have extracted conformal data from the four-point function $G_0(x,\bar x)$, parameterized by the covering-surface coordinate $x$, and obtained in Eq.(\ref{r0fun}).
It is instructive to compare this function with its R-charged counterpart $G_+(x,\bar x)$, given in Eq.(\ref{GRchar}). Although the R-neutral and the R-charged fields have the same twist and the same dimensions, the corresponding four-point functions have different analytic structures.
For generic $n$, the positions of the branching points of $G_+$ and $G_0$ coincide (for special values of $n$, in each case, the branching points may become non-branched zeros or poles);
meanwhile, $G_0$ has two additional simple zeros at $x = \frac{1-n}{2} \pm \frac{\sqrt{n^2-1}}{2}$.
In particular, the two functions have singularities at the same points, $x_{\frak a} = \{ -n, \frac{1-n}{2}, 0 , \infty\}$,%
\footnote{%
Note that these points have negative exponents for all $n > 2$.
}
with the branching structure at these points being different in each case.
This is, of course, a direct consequence of the covering map parameterization, and of the fact that the singular points correspond to (non-trivial) OPE limits: at $x_{\frak a} = \{\frac{n-1}{2} , \infty\}$, both $G_+$ and $G_0$ have the same behavior, because this is the coincidence limit of the interaction operators present in both functions, while at $x_{\frak a} = \{ 0, - n\}$ the functions have different branching structures,  reflecting the different dynamics of the R-charged and R-neutral fields as they are taken near the interaction operator.

In recent works \cite{Lima:2020boh,Lima:2020kek}, we have shown that the R-charged Ramond ground states in the $n$-twisted sectors,  $R^{\pm}_{[n]}$, acquire $\la^2$-dependent anomalous dimensions in perturbation theory, for $n < N$.
For $n = N$,  the dimensions are protected at least at leading order in the genus-zero, large-$N$ approximation.
Here we have expanded these results to include the remaining Ramond fields $R^{0\pm}_{[n]}$, with zero R-charge.
Although the dynamics is different, the results are qualitatively similar: Ramond fields with non-maximal twists are renormalized, and their dimensions are lifted at order $\la^2$.
The qualitative agreement of behavior between the renormalization properties of R-charged and R-neutral Ramond fields extends beyond the results presented here to include composite fields \cite{Lima:2020nnx}; in particular, there is no renormalization of the Ramond ground states of the full orbifold theory, i.e.~$\prod_i (R^{(i)}_{n_i})^{k_i}$,
with 
$\sum_i k_i n_i = N$,
where each $R^{(i)}_{n_i}$ is an R-neutral or an R-charged Ramond field \cite{Lima:2021wrz}.

Let us note that using Ward identities, it is possible to use the functions computed here and in 
\cite{Lima:2020boh,Lima:2020kek,Lima:2020nnx,Lima:2021wrz}
to obtain correlators involving excited fields such as the ones relevant for the study of the three charge D1-D5-P system \cite{Giusto:2004id,Giusto:2004ip,Giusto:2004kj,Giusto:2012yz,Giusto:2015dfa,Bena:2017xbt,Hampton:2018ygz}, in a rather straightforward manner.
In contrast to the 1/4-BPS two-charge solutions, 
the R-neutral Ramond ground states do enter as crucial ingredients of the CFT realization of 1/8-BPS microstates where, instead of the product of pure Ramond fields $\prod_i (R^{(i)}_{n_i})^{k_i}$, now the R-neutral fields must be excited by R-current modes, for example  $[R^+_1]^{N - kn} [ (J^+_{-1})^{\ell} R^{0\pm}_n ]^k$ see e.g.~\cite{Giusto:2015dfa}.
More than being tools for obtaining anomalous dimensions, the four-point functions we have computed contain important information about the dynamics of the Ramond fields, and the effect of the interaction operator upon them.
We found new non-BPS fields $\cal Y^{0\pm}_{[m]}$ of conformal weight $h^{\cal Y}_m = \frac{m}{4} + \frac{3}{4m}$ in the OPE channels, with twists $m$  dictated  by the composition of the cycles of  $R^{0\pm}_{[n]}$ and $\Oint$.
This R-neutral doublet should be compared with the corresponding R-charged doublet $Y^{\pm}_{[m]}$ found in the OPEs with the R-charged Ramond fields $R^{\pm}_{[n]}$ and $\Oint$,  whose conformal weight is $h^Y_m = \frac{m}{4} + \frac{5}{4m}$ \cite{Lima:2020kek}.
The OPEs and the conformal blocks are a fundamental part of a conformal field theory, and since the twisted Ramond fields are very basic building blocks of the orbifold CFT, exploring their detailed dynamics as one moves the theory away from the free orbifold point is an important task.
In this respect, we would like to investigate further the nature of the ``intermediate'' fields ${\cal Y}^{0\pm}_{[m]}$, their relation to the fields ${\cal A}^{0\pm}_{[m]}$ and ${\cal B}^{0\pm}_{[m]}$ appearing in the OPE of the interaction operator with $O^{(1,1)\pm}_{[n]}$, and how these OPEs are related to the failing of the Ramond fields from being protected, while the chiral fields are.

\bigskip

\noindent
{\bf Acknowledgements}

\noindent
The work of M.S. is partially supported by the Bulgarian NSF grant KP-06-H28/5 and that of M.S. and G.M.S. by the Bulgarian NSF grant KP-06-H38/11.
 M.S. is grateful for the kind hospitality of the Federal University of Esp\'irito Santo, Vit\'oria, Brazil, where part of his work was done.

\appendix

\section{Four-point functions with R-charged fields}	\label{SectLMRcharged}

Let us use the LM technique to compute the four-point function
\be\label{gplus}
G_+(u,\bar u) = 
	\Big\langle 
	R^{-}_{[n]}(\infty) \Oint (1) \Oint (u,\bar u) R^{+}_{[n]}(0) 
	\Big\rangle
\ee
for the R-charged Ramond fields $R^{\pm 0}_{[n]}$. 
This function was found in \cite{Lima:2020boh,Lima:2020kek} via the stress-tensor method. 
The goal here is to show that it is much simpler to find than the R-neutral function of Sect.\ref{SectLMtech}.
Since the Ramond fields are purely fermionic, the bosonic factor $G_B$ of the correlator is the same as before, given by Eq.(\ref{Gbosn}). 
The twist factor $G_\s = e^{S_L}$ is universal for the fixed twist structure, and again given by the Liouville action (\ref{Liouvx}).
What we need to compute anew is the fermionic factor $G_F$.
The R-charged Ramond fields lift to 
\be	\label{mapr}
\begin{split}
R_n^{\pm} (0,\bar 0) 
	&\mapsfrom b_{0}^{-{1\over 4n}}
			e^{\pm {i\over 2}(\phi^1+\phi^2)}(0)\times c.c
\\
R_n^{\pm} (\infty,\bar \infty) 
	&\mapsfrom b_{\infty}^{{1\over 4n}}
			e^{\pm {i\over 2}(\phi^1+\phi^2)}(\infty)\times c.c
\end{split}
\ee
Let us first consider the first term $I$ in the interaction product (\ref{ints}).
The holomorphic correlator is, now, instead of (\ref{Expsnuethol}), 
\be	\label{expCorh}
\Big\langle 
	e^{-{i\over 2}(\phi^1+\phi^2)}(\infty)
	e^{-{i\over 2}(\phi^1-\phi^2)}(t_1)
	e^{{i\over 2}(\phi^1-\phi^2)}(x)
	e^{{i\over 2}(\phi^1+\phi^2)}(0)
	\Big\rangle
=	
	(t_1-x)^{-{1\over 2}} ,
\ee
and the anti-holomorphic part is, instead of (\ref{ExpsnuetAnhol}),
\be	\label{expCorah}
\begin{split}
\Big\langle 
e^{{i\over 2}(\bphi^1 + \bphi^2)}(\bar\infty)
&	\Big[
	 e^{{i\over 2}(\bphi^1-\bphi^2 )}(\bar t_1)
	 e^{-{i\over 2}(\bphi^1-\bphi^2)}(\bar x)
\\
&
\qquad\qquad
	 +
	 e^{-{i\over 2}(\bphi^1-\bphi^2)}(\bar t_1)
	 e^{{i\over 2}(\bphi^1-\bphi^2)}(\bar x) 
	 \Big]
e^{-{i\over 2}(\bphi^1+\bphi^2)}(\bar 0)
\Big\rangle
\\
&
=
	2 (\bar t_1-\bar x)^{-{1\over 2}}
\end{split}
\ee
Next, it is easy to see that all the other interaction terms $II,III,IV$ give the exactly the same result.
The product of the two correlators (\ref{expCorh}) and (\ref{expCorah}) gives, directly, a real function $|t_1 - x|^{-1}$.
By contrast, the product of correlators (\ref{Expsnuethol}) and (\ref{ExpsnuetAnhol}) is \emph{not} real, only the \emph{sum} of the four terms $I,II,III,IV$ is real.

The fermionic factor takes into account also the factors coming from the covering map,
\be
G_F = 	\Big|  
	b_0^{-{1\over 4n}} 
	b_\infty^{{1\over 4n}} 
	b_1^{-{1\over 8}} 
	b_x^{-{1\over 8}} 
	(t_1-x)^{-\frac12}
		\Big|^2
\ee
Multiplying $G_\s \times G_B \times G_F$ by the additional factor (\ref{addbfac}), expressing everything explicitly in terms of $x$ using (\ref{biss}), we obtain
\be	\label{GRchar}
G_+(x,\bar x) = 
	\left|
	C
	\frac{
	x^{-{5n\over 4}+{5\over 2}}
	(x-1)^{{5n\over 4}+{5\over 2}}
	(x+n)^{-{3n\over 4}+{1\over 2}}
	(x+n-1)^{{3n\over 4}+{1\over 2}}
	}{
	(x+{n-1\over 2})^4
	}
	\right|^2 ,
\ee
in agreement with \cite{Lima:2020boh,Lima:2020kek}.

\bibliographystyle{utphys}

\bibliography{RNeutralShortPaperReferencesV29c} 

\end{document}